\documentclass[twocolumn]{autart}
\usepackage[english]{babel}

\usepackage{amsmath, amssymb}
\usepackage{enumerate}
\usepackage{graphicx}  
\usepackage[dvipsnames]{xcolor}
\graphicspath{{figs/}}  

\usepackage{subcaption}  

\usepackage{algorithm}
\usepackage{algpseudocode}
\algrenewcommand\algorithmicrequire{\textbf{Input:}}
\algrenewcommand\algorithmicensure{\textbf{Output:}}

\usepackage[overload]{empheq}
\usepackage{cite}

\usepackage[hyperindex=true, breaklinks=true,colorlinks=true,bookmarks=false, pdftitle={Dissipativity and Turnpike Properties in Generalized Nash Equilibrium Problems},pdfauthor={Authors}, colorlinks=true, pagebackref=false, plainpages=false, pdfpagelabels, linkcolor=black, citecolor=black, filecolor=black, urlcolor=black]{hyperref}

\definecolor{lightgray}{gray}{0.9}
\definecolor{ggreen}{rgb}{0,0.5,0}
\definecolor{rred}{rgb}{0.5,0,0}

\newcommand{\mbf}[1]{\mathbf{#1}}
\newcommand{\bb}[1]{\mathbb{#1}}

\newcommand{\mc}[1]{\mathcal{#1}}

\newcommand{\R}{\mathbb{R}}
\newcommand{\x}{\mathbf{x}}

\DeclareMathOperator*{\argmin}{argmin}

\DeclareMathOperator{\subjectto}{\ {s.}{t.}\ }

\newtheorem{ass}{Assumption}
\newtheorem{dfn}{Definition}

\newtheorem{prp}{Proposition}
\newtheorem{rmk}{Remark}
\newtheorem{ex}{Example}

\newenvironment{proof}{\paragraph*{Proof}}{\hfill$\square$}

\newcommand{\sh}[1]{\textcolor{NavyBlue}{[#1]\raise 0.5ex \hbox{\footnotesize{SH}}}}
\newcommand{\tf}[1]{\textcolor{Orange}{[#1]\raise 0.5ex \hbox{\footnotesize{TF}}}}

\begin{document}

\begin{frontmatter}

\title{System-Theoretic Analysis of Dynamic Generalized Nash Equilibria -- Turnpikes and Dissipativity\thanksref{footnoteinfo}}

\author[ETH]{Sophie Hall}\ead{shall@ethz.ch},    
\author[ETH]{Florian D\"orfler}\ead{dorfler@ethz.ch},               
\author[TUH]{Timm Faulwasser}\ead{timm.faulwasser@ieee.org}  

\address[ETH]{Automatic Control Laboratory, ETH Z\"urich, Switzerland}
\address[TUH]{Institute of Control Systems, Hamburg University of Technology, Germany}            
\thanks[footnoteinfo]{The authors are with the  Automatic Control Lab, ETH Z\"urich, Physikstrasse 3, 8092 Z\"urich, Switzerland and with the Institute of Control Systems, Hamburg University of Technology, Harburger Schlo{\ss}stra{\ss}e 22a, 21079 Hamburg, Germany. This work was supported by the Swiss National Science Foundation under the NCCR Automation (grant 51NF40 225155).}
        
\begin{keyword}                           
dynamic games, optimal control, generalized Nash equilibrium problems, multi-agent systems, dissipativity, turnpike property, non-cooperative game theory          
\end{keyword}                       

\begin{abstract} Generalized Nash equilibria are used in multi-agent control applications to model strategic interactions between agents that are coupled in the cost, dynamics, and constraints, and provide the foundations for game-theoretic MPC (Receding Horizon Games). 
We study properties of finite-horizon dynamic GNE trajectories from a system-theoretic perspective. We show how strict dissipativity generates the turnpike phenomenon in GNE solutions. Moreover, we establish a converse turnpike result, i.e., the implication from turnpike to strict dissipativity. 
We derive conditions under which the steady-state GNE is the optimal operating point and, using a game value function, we give a local characterization of the geometry of storage functions. Finally, we design linear terminal penalties that ensure dynamic GNE trajectories applied in open-loop converge to and remain at the steady-state GNE. These connections provide the foundation for future system-theoretic analysis of GNEs similar to those existing in optimal control as well as for recursive feasibility and closed-loop stability results of game-theoretic MPC.
\end{abstract}
\end{frontmatter}

\section{Introduction}


Control is at the core of many multi-agent applications managing interactions between strategic agents with conflicting objectives, dynamics, and coupled action spaces. Generalized Nash equilibria (GNEs) have emerged as a promising solution concept at the intersection of game theory and optimal control~\cite{facchinei2009generalized}. They naturally enable to define per-agent optimal control problems that are coupled, in a game-theoretic sense, in the objective, constraints, and dynamics. GNEs have been applied to resource allocation problems in dynamic settings (e.g., energy~\cite{atzeni2013demand, hall2022receding}, transportation~\cite{bassanini2002allocation}, telecommunications~\cite{pavel2012game}) and competitive dynamic settings (e.g., autonomous driving~\cite{dreves2018generalized, lecleach2022algames}, supply chains~\cite{hall2024receding}).


Applications in real-time control of dynamic multi-agent systems catalyzed extensive research on efficient GNE-seeking algorithms and their convergence guarantees~\cite{gadjov2019passivity, belgioioso2022distributed}. The resulting solution sets have been studied in detail~\cite{kulkarni2012variational, nabetani2009parametrized},  as well as their efficiency~\cite{kulkarni2019efficiency} and fairness properties~\cite{hall2025limits}.

A fundamental step towards safe deployment of controllers is the characterization of their closed-loop properties.  Yet, the behavior of finite-horizon trajectories resulting from GNEs remains poorly understood, even in the open-loop case. Previous works have applied dissipativity techniques to analyze the feedback mapping~\cite{hall2025stability} and drawn connections to dynamic variational inequalities~\cite{benenati2025linear}. Despite this progress, finite-horizon GNE trajectories lack the rich system-theoretic characterizations developed for optimal control problems over the past decades, cf.~\cite{Willems71a,Anderson87a,Trelat15a} and many others.


\begin{figure}[h]
        \centering
        \includegraphics[width=\columnwidth]{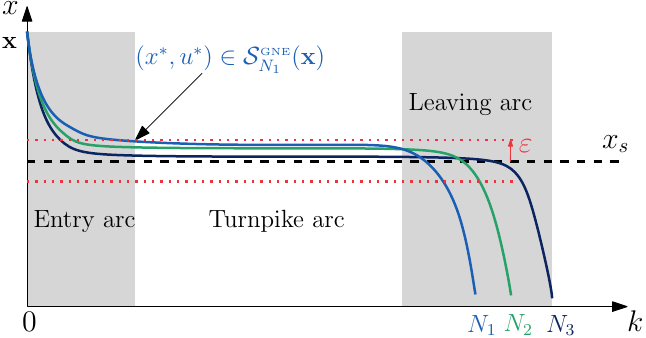}
        \caption{Schematic of a dynamic GNE state trajectory $x^*$ exhibiting the turnpike property at the steady-state GNE $x_s$ for different horizon lengths.\label{fig:Turnpike_Schematic}}
  \end{figure}   
In this context, the turnpike phenomenon arises as a property of solutions to an optimal control problem (OCP) solved for different initial conditions and varying horizon lengths. The phenomenon is characterized by the solutions clustering near one common steady state which is referred to as the \text{turnpike}, see Figure~\ref{fig:Turnpike_Schematic}. It is a similarity property of parametric OCPs and thus is a crucial element in the closed-loop analysis of OCPs applied in a receding-horizon fashion, i.e., in model predictive control (MPC).
First observations date back to~\cite{Ramsey28} and~\cite{vonNeumann38}, while the notion \textit{turnpike} was coined by~\cite{Dorfman58}. We refer to~\cite{Mckenzie76} and~\cite{faulwasser2022turnpike} for classic and modern literature overviews. A recent key development is the uncovering of the close connection between the turnpike phenomenon and dissipativity properties~\cite{gruene2016relation,epfl:faulwasser15h}, which is in turn linked to the dissipativity route to infinite-horizon optimal control which originated with~\cite{Willems71a},  see ~\cite{tudo:faulwasser21a} for recent nonlinear extensions. The link between turnpike and dissipativity enabled major breakthroughs in the closed-loop stability analysis of nonlinear and economic MPC, see, e.g., \cite{gruene2013economic}. 
In applications such as energy management, the turnpike effect or end-of-horizon effect leads, for instance, to myopic discharging of batteries~\cite{hall2022receding,vincent2020influence}.


Interestingly, turnpike phenomena have also been observed in games in the Economics literature as early as the 1980s~\cite{fershtman1986turnpike} and were later treated from a theoretical perspective for infinite-horizon open-loop games in continuous time~\cite{carlson1995turnpike} and discrete time~\cite{carlson1996turnpike} focusing on existence, uniqueness of equilibria and convergence to the turnpike. Extensions to differential games with coupled state constraints followed~\cite{carlson2000infinite}. Turnpike properties in games have recently received renewed attention:
The specialized class of stochastic differential LQ games with continuous-time dynamics are studied in~\cite{li2025turnpike} for the zero-sum setting and in~\cite{cohen2025turnpike} for the non-zero sum setting. Multiple works derive turnpike properties for mean field games under a large population assumption~\cite{cirant2021long, carmona2024leveraging, ersland2025long, fedorov2025studying}. Further, turnpikes were observed in competitive dynamic supply chains~\cite{hall2024receding}.



This paper lays the foundations for an overarching theory of turnpike and dissipativity analysis in noncooperative games for a general class of cost functions, constraints, and discrete-time dynamics. We build a bridge between the dissipativity route to turnpikes in optimal control and dynamic games. Specifically, we establish the turnpike property for finite-horizon discrete-time dynamic GNE problems and study its implications on dissipativity of the GNE with respect to the steady-state GNE. Overall, we make the following four contributions:

\begin{enumerate}
    \item[1.] We demonstrate the crucial structural link between turnpike properties in parametric OCPs and parametric GNEs providing a foundation for system-theoretic analysis of GNEs.  
    \item[2.] We establish that strict dissipativity implies the turnpike property, also in GNEs. Further, we derive a converse turnpike result, connecting the turnpike of the open-loop trajectory to dissipativity with respect to the steady-state GNE.  
    \item[3.] We provide an optimality-like interpretation of GNEs by considering a game value function and studying its local gradient structure. Additionally, we show that the storage function's gradient equals the sum of agents' dual multipliers at the steady-state GNE.
    \item[4.] We design mechanisms to suppress the leaving arc in GNE open-loop trajectories, using per-agent linear terminal penalties to ensure trajectories converge to and remain at the steady-state GNE.
\end{enumerate}


The remainder of this paper is organized as follows. Section~\ref{sec:problem} formulates the finite-horizon GNE problem for which we establish turnpike and dissipativity results in Section~\ref{sec:DissTurn}. We study connections between KKT systems of OCPs and GNEs in Section~\ref{sec:OCP-GNE-relation} and conclude the paper with simulation results in Section~\ref{sec:Sim}.


\subsection*{Notation} We denote by $\bb{Z}_K= \{0,\dots, K\!-\!1\}$ the sequence of the first $K$ non-negative integers. Given $M$ vectors $u^1, ..., u^M$, we denote  by  $ u = \text{col}(u^v)_{v=1}^M:= [(u^1)^\top, \ldots, (u^M)^\top]^\top$ the stacked vector of vectors $u^v$, where $u^v$ is the decision vector of agent $v$, and of all other agents as $u^{-v} = \text{col}(\{u^{s}\}_{s\in \mc{M}\backslash v})$. Our use of class $\mc{K}$, $\mc{KL}$, and $\mc{K}_\infty$ comparison functions follows standard conventions~\cite{Kellett14}. For a finite set $\mc{Q}$, we denote its cardinality by $\#\mc{Q}$. Let $\mc B_\varepsilon(\x) \subset \mathbb R^n$ denote an open ball of radius $\varepsilon$, centered at $\x$.


\section{Finite-Horizon Generalized Nash Equilibrium Problems}\label{sec:problem}
Consider a nonlinear discrete-time system 
\[
x_{k+1} =  f(x_k,u_k), \quad x_0 = \mathbf x
\]
with state $x_k \in \mathbb{R}^{n_x}$, control input $u_k \in \mathbb{R}^{n_u}$, and initial condition $\x$ which lies in a compact set $\x \in \bb{X}_0$.
Suppose the input is not computed by a central entity but is instead a stacked vector $u_k = [(u_k^1)^\top, \ldots, (u_k^M)^\top]^\top$, where each component $u_k^v$ is decided by a self-interested agent $v\in \mc{V} := \{1,\dots, M\}$. Each agent aims to steer the shared dynamics in its own favor
\begin{equation}\label{eq:Dynamics}
x_{k+1} =  f(x_k,u_k^v,u_k^{-v}), \quad x_0 = \mathbf x.
\end{equation}
%
%
%
Specifically, each agent $v$ minimizes its accumulated stage cost $\ell^v$ over a prediction horizon $N$ subject to the shared dynamics~\eqref{eq:Dynamics}. In addition to this dynamic coupling, agents influence each other's outcomes through coupling in their cost and constraints, while still being subject to individual local constraints. Formalizing this gives rise to a finite-horizon dynamic Generalized Nash Equilibrium Problem (GNEP) of the form
\begin{subequations}
\label{eq:MPCPerAgent}
\begin{empheq}[left=\forall v\in \mc{V}:  \empheqlbrace]{align}
\label{eq:RunningCost}
\displaystyle \min_{u^v,\, x}  &\;\sum_{k= 0}^{N-1} \ell^v(x_k, u_k^v,u_k^{-v})\\
\textrm{s.t.} \quad &  x_{k+1} =  f(x_k,u_k^v,u_k^{-v})  \hspace{1.25em}  k \in \bb{Z}_{N} \label{eq:Constr1}\\
&g(x_k,u_k^v, u^{-v}_k) \leq 0, \hspace{2.5em} k \in \bb{Z}_{N} \label{eq:Constr2}\\
  & h^v(u_k^v)\leq 0,      \hspace{5.6em} k \in \bb{Z}_{N} \label{eq:Constr3}\\ 
&\; x_0 = \mbf{x}, \label{eq:Constr4}
\end{empheq}
\end{subequations}
where~\eqref{eq:Constr3} are nonlinear local constraints and~\eqref{eq:Constr2} nonlinear coupled constraints. 

For each choice of the initial condition $\x$ and the horizon $N$, the GNEP~\eqref{eq:MPCPerAgent} generates \textit{finite-horizon GNE trajectories}. Hence, we view the  GNEP~\eqref{eq:MPCPerAgent} as a parametric game-theoretic decision problem. Note that---for now---we make no assumptions on the functions in~\eqref{eq:MPCPerAgent} but will add them when our theoretical results require them.

We define the following per-agent and global action sets
\begin{align}
&\mc{Z}_N^v( \x, u^{-v}) = \{(x,u^v)  ~|~ \eqref{eq:Constr1} - \eqref{eq:Constr4}\} \\[5pt]
&\mc{Z}_N(\x) = \{(x, u) \in \R^{(N+1)n_x+N n_u}~|~  \eqref{eq:Constr1} - \eqref{eq:Constr4}\} \label{eq:FeasibleSet}
\end{align}
as well as $\mc{Z}_{\infty}(\x)$ for the infinite horizon problem and the per-agent cumulative cost 
\[
J_N^v(x,u^v, u^{-v}) :=\sum_{k= 0}^{N-1} \ell^v(x_k, u_k^v, u^{-v}_k).
\]
With these abstractions we can write~\eqref{eq:MPCPerAgent} as follows:
\begin{equation}\label{eq:GNEPcompact}
\forall v\in \mc{V}: \; 
\left\{\begin{aligned}\;\min_{u^v, x}\;& J^v_N(x,u^{v},u^{-v}) \\
\subjectto &(x,u^{v})\in \mc{Z}_N^v( \x, u^{-v}).
\end{aligned}\right.
\end{equation}
%
Decisions that jointly solve~\eqref{eq:MPCPerAgent} are called \textit{generalized Nash equilibria} (GNE)~\cite[\S 2]{facchinei2009nash}. Intuitively, at a GNE no agent $v \in \mc{V}$ can reduce its cost by unilaterally changing its own decision.
\begin{dfn}[Generalized Nash equilibrium] \label{dfn:GNE} ~\\A joint decision $(x^*, u^*)\in \mc{Z}_N(\x)$ is a GNE  of~\eqref{eq:MPCPerAgent} if 
\begin{align*}
\forall v\in \mc{V}: \, J_N^v(x^{*}, u^{v*}, u^{-v*} ) \leq J_N^v(x, u^{v}, u^{-v*}) 
\end{align*} $ \forall (x,u^v) \in \mc{Z}_N( \x, u^{-v*})$. The corresponding solution set for fixed $N\in \bb{N}$ and $\x\in \bb{X}_0$ is denoted as
\begin{align}
(x^*,u^*) \in \mc{S}^{\text{\tiny GNE}}_N(\x) \subset \R^{(N+1)n_x + N n_u}.
\end{align}
\end{dfn}
In analogy to optimal control concepts, we refer to $(x^*,u^*) \in \mc{S}^{\text{\tiny GNE}}_N(\x) $ as a \textit{GNE pair} or an \textit{optimal game pair}. As a shorthand, the point-wise in time projection of $\mc{S}^{\text{\tiny GNE}}_N(\x)$ onto the states is written as 
$\mathbb X_N(\bb{X}_0)\subseteq \R^{n_x}$.
That is, if $\tilde x \in \mathbb X_N(\bb{X}_0)$ then there exist at least one game pair $(x^*,u^*) \in \mc{S}^{\text{\tiny GNE}}_N(\x), \x \in \bb X_0$ such that the state trajectory $x^*$ passes through $\tilde x$ at least once. Similarly $X_{\infty}(\bb{X}_0)$ refers to the infinite horizon problem.

Henceforth, we study convergence or clustering of finite-horizon GNE trajectories close to a two-fold equilibrium: a point that is both (i) a steady-state of~\eqref{eq:Dynamics}, i.e., $\bar x = f(\bar x, \bar u^v, \bar u^{-v})$; and (ii) a strategic (decision) equilibrium of the one-step GNEP in~\eqref{eq:MPCPerAgent} such that no agent benefits from deviating in the next time step. 
This is formalized in the following \textit{steady-state GNEP}.
\begin{dfn}[Steady-state GNE] The pair $(x_s,u_s)$ is called a steady-state GNE if it solves 
\begin{align}\label{eq:SteadyStateGNEP}
v\in \mc{V}: \left\{
\begin{array}{r l}
\displaystyle \min_{\bar{u}^v, \bar{x}} & \; \ell^v(\bar{x}, \bar{u}^v, \bar{u}^{-v}) \\ 
 \subjectto  &  f(\bar{x},\bar{u}^v,\bar{u}^{-v})- \bar{x}=0\\ 
            &  g(\bar{x},\bar{u}^v, \bar{u}^{-v}) \leq 0,\\
            &h^v(\bar u^v)\leq 0,\\
\end{array} 
\right.
\end{align}
with the corresponding solution set $\mc{S}^{\text{\tiny GNE}}_s \subset \R^{n_x + n_u}.$ Problem \eqref{eq:SteadyStateGNEP} is called a steady-state GNEP.
\end{dfn}
We analyze properties of the GNE in~\eqref{eq:MPCPerAgent} and \eqref{eq:SteadyStateGNEP} provided a solution exists. Thus, in the following we assume that  $\mc{S}^{\text{\tiny GNE}}_N(\x)$ and $\mc{S}^{\text{\tiny GNE}}_s $ are nonempty $\forall \x \in \bb{X}_0$.

\begin{rmk}[Existence and computation]
The theoretical results which follow hold for the nonlinear GNE problem in~\eqref{eq:MPCPerAgent}. Existence of a GNE has been established under a variety of assumptions, as summarized in~\cite{dutang2013existence}. One approach is to assume that $\forall v\in \mc{V}, \; \forall \x \in\mathbb{X}_0,\;  \forall u^{-v}\in \mc{U}^{-v}:= \Pi_{j\in\mc{V}\backslash v} \mc{U}^j \ $ the set $\mc{Z}_N^v( \x, u^{-v})$ and $\mc{U}^v$ is nonempty, compact and convex and the function  $J^v_N(\cdot, \cdot, u^{-v})$ is convex in both arguments~\cite[Prop. 12.7]{palomar2010convex}.  
Efficient algorithms to compute solutions of nonlinear GNEPs are a topic of ongoing research, one example is given in~\cite{lecleach2022algames}. 
\end{rmk}
\section{Dissipativity and Turnpikes 
for GNEPs}\label{sec:DissTurn}
Turnpike and dissipativity properties are closely linked, providing system-theoretic insights for optimal control. In the following, we demonstrate that this link also exists for the GNE problem in~\eqref{eq:MPCPerAgent}. 

Consider the social welfare as a joint performance measure for the population of agents
\begin{equation} \label{eq:ellStack}
    \ell(x_k,u_k) := \sum_{v\in \mathcal V} \ell^v(x_k,u_k^v, u_k^{-v})
\end{equation}
and introduce the shorthand
\begin{equation}\label{eq:overallJ}
  J_N(x,u) :=\sum_{k= 0}^{N-1} \ell(x_k, u_k).
\end{equation}

\begin{rmk}[GNEPs vs. OCPs]
    ~\\At this point it is fair to ask for the precise distinction between the GNE setting considered here and an optimal control point of view. Solving~\eqref{eq:GNEPcompact} yields a GNE, i.e., a joint decision from which no agent $v \in \mc{V}$ can reduce its cost unilaterally. In contrast, the optimal control counterpart of~\eqref{eq:GNEPcompact} reads
    \begin{align*}\;
        V^\diamond_N(\x) := \min_{u,x}\; J_N(x,u) 
        \subjectto (x,u)\in \mc{Z}_N(\x, u),
    \end{align*}
    where all control actions are chosen with the sole objective of minimizing the cost for the entire agent population~\eqref{eq:overallJ}. Whenever necessary, we use superscript $\cdot^*$ for GNEP quantities and $\cdot^\diamond$ for OCP quantities.
    
    Take a simple unconstrained quadratic example with two agents and cost $J^1(u^1,u^2) = r^{1,1} (u^1)^2 + r^{1,2} u^1u^2 + r^1_{\text{lin}} u^1$, and analogously for agent~2. Any equilibrium solution must satisfy \[ \left[  \begin{smallmatrix} \frac{\partial J^1(u^1,u^2) }{\partial u^1}\\  \frac{\partial J^2(u^1,u^2) }{\partial u^2}\end{smallmatrix}\right]  =    \underbrace{\left[\begin{smallmatrix} 2r^{1,1} & r^{1,2} \\  r^{2,1} & 2r^{2,2}\end{smallmatrix}\right]}_{R^*} u +  \left[  \begin{smallmatrix}r^1_{\text{lin}} \\  r^2_{\text{lin}}\end{smallmatrix}\right] \stackrel{!}{=} 0 \]
    whereas $R^\diamond= \left[\begin{smallmatrix} 2r^{1,1} & 0.5 (r^{1,2}+ r^{2,1}) \\  0.5 (r^{1,2}+ r^{2,1}) & 2r^{2,2}\end{smallmatrix}\right]$ and thus $u^*$ and $u^\diamond$ diverge with diverging $r^{1,2}$ and $r^{2,1}$. It is well known that in noncooperative games where agents act selfishly, the equilibrium solution does not minimize the sum of costs, and thus differs from the OCP solution. This maximal efficiency loss is referred to as the \textit{price of anarchy} (PoA); see~\cite{kulkarni2019efficiency} for an analysis in the context of GNEs. Formally,
    \begin{equation}\label{eq:PoA}
        \text{PoA}(\x) :=  \dfrac{\displaystyle \sup_{(x^*, u^*) \in \mc{S}^{\text{\tiny GNE}}_N(\x)} J_N(x^*, u^*)}{V^\diamond_N(\mathbf{x})} \geq 1.
    \end{equation}
\end{rmk}


Next, using the joint performance measure $J_N$ in ~\eqref{eq:overallJ}, we introduce a strict dissipativity notion which follows concepts used in optimal control.
 \begin{dfn}[Strict dissipativity of GNEPs]\label{dfn:StrictDiss}
Given a steady-state GNE $(x_s,u_s)\in \mc{S}^{\text{\tiny GNE}}_s$, the GNEP~\eqref{eq:MPCPerAgent} is called strictly dissipative with supply rate 
\[s(x_k,u_k) := \ell(x_k,u_k) -\ell(x_s,u_s)\]
if
there exists a storage function $\Lambda: \bb X_N(\bb{X}_0)\to \R$ bounded from below  such that $\forall N \in \bb{N}, \forall \x \in \bb{X}_0$
 \begin{multline}\label{eq:GsDI}
\Lambda(f(x_k,u_k)) -\Lambda(x_k) \leq \\ -  \alpha_{\ell}\left(\left\|
\begin{matrix}x_k- x_s\\ u_k - u_s \end{matrix}
\right\|\right) +s(x_k,u_k) \tag{sDI}
 \end{multline}
holds for some $\alpha_{\ell}\in \mc{K}$ and each point $(x_k,u_k)$ along game pairs $(x,u)\in \mc{S}^{\text{\tiny GNE}}_N(\x)$. If \eqref{eq:GsDI} holds for $\alpha_{\ell} \equiv 0$, the GNEP~\eqref{eq:MPCPerAgent} 
is called dissipative. 
 \end{dfn}
Since $\alpha_\ell$ is of class $\mathcal{K}$, inequality~\eqref{eq:GsDI} is strict whenever $(x_k, u_k) \neq (x_s, u_s)$, hence the term \textit{strict dissipativity} is used in the literature. A natural question is how to verify strict dissipativity of GNEPs. To this end, consider

%

\begin{equation}\label{eq:AvailStorDef}
   \Lambda_{\alpha_\ell}(\x) := \displaystyle \hspace{-2mm}\sup_{\substack{N\in \bb{N}\\ (x^*,u^*)\\  \in\mc{S}^{\text{\tiny GNE}}_N(\x)}} \hspace{-1mm}\sum_{k=0}^{N-1} \alpha_{\ell}\left(\left\| \begin{smallmatrix}x_k^*- x_s\\ u_k^* - u_s \end{smallmatrix}\right\|\right) - s(x_k^*, u_k^*).
\end{equation}
Observe that, given the initial condition $\x$, this is a free end-time optimization problem over all game pairs $(x^*,u^*) \in\mc{S}^{\text{\tiny GNE}}_N(\x)$. The objective is to compute the maximum storage that can be extracted from the system, hence $\Lambda_{\alpha_{\ell}}$ is referred to as the \textit{available storage} \cite{Willems72a,byrnes1994losslessness}. The next result extends the classical characterization of dissipativity via available storage~\cite{Willems72a} to dynamic GNEPs.
\begin{thm}[Available storage of GNEPs \& \eqref{eq:GsDI}]\label{thm:AvailStor}
The GNEP~\eqref{eq:MPCPerAgent} is strictly dissipative with supply rate $s(x_k,u_k) = \ell(x_k,u_k) - \ell(x_s,u_s)$ and $\alpha_{\ell}\in \mc{K}$ (Definition \ref{dfn:StrictDiss}) if and only if, for all $\x \in \bb X_\infty(\bb{X}_0)$, the bound
%
    $\displaystyle \Lambda_{\alpha_\ell}(\x) <\infty$ holds.
\end{thm}
\vspace{-\baselineskip}
\begin{proof} The proof is given in Appendix~\ref{apendix:ProofThmAvailStor}.
\end{proof}


Willems' motivation for the introduction of dissipativity notions in systems and control has been the \textit{``generalization of Lyapunov functions to open systems, to systems with inputs and outputs''}~\cite{Willems07a}. From this perspective, the storage function is the amount of energy in an open system and the available storage the amount that can be extracted via inputs and outputs. Thus, Theorem~\ref{thm:AvailStor} states that strict dissipativity, which loosely implies that solutions are attracted to $(x^s, u^s)$, can only hold if for all initial conditions the energy extractable along GNE trajectories is bounded; if it were unbounded, no such attraction could occur.
%

In OCPs, strict dissipativity is often assumed to hold along all feasible trajectories~\cite[Defn. 2.1]{gruene2016relation}, though~\cite{epfl:faulwasser15h} restrict it to optimal ones. As the next result shows, in the game-theoretic setting the restriction to GNE trajectories of~\eqref{eq:MPCPerAgent} is indeed necessary.



\begin{lem}(Restriction of domain of~\eqref{eq:GsDI}) The strict dissipativity inequality~\eqref{eq:GsDI} does not necessarily hold along all feasible trajectories $(x,u)\in \mc{Z}_N(\x)$.
\end{lem}%
\vspace{-\baselineskip}
\begin{proof} We introduce the optimal control counterpart of the steady-state game in~\eqref{eq:SteadyStateGNEP} which gives the steady-state minimizer
\begin{align}\label{eq:OCPsteadystate}
(x^\diamond, u^\diamond) = &\argmin_{\bar x, \bar u} \; \ell(\bar x, \bar u) \; \\\nonumber
&\text{s.t.}\;  \bar x = f(\bar x,\bar u), \; g(\bar{x},\bar{u}) \leq 0,\;  h(\bar u) \leq 0.
\end{align} 
in which $\ell(\bar x, \bar u) = \displaystyle \sum_{v\in \mc{V}}\ell^v(\bar x, \bar u^v,u^{-v})$ as defined in~\eqref{eq:ellStack} and $h(\cdot)$ stacks all local constraints.
Consider the dissipativity inequality in~\eqref{eq:GsDI} 
 \begin{multline*}
\Lambda(f(x_k,u_k)) -\Lambda(x_k) \leq 
s(x_k,u_k)-  \alpha_{\ell}\left(\left\|
\begin{smallmatrix}x_k- x_s\\ u_k - u_s \end{smallmatrix}
\right\|\right)  
 \end{multline*} 
 which evaluated at $(x^\diamond, u^\diamond) $ for $s(x_k,u_k) = \ell(x_k,u_k)-\ell(x_s, u_s)$ as given in Definition~\ref{dfn:StrictDiss} yields 
\begin{align} \label{eq:DissAtOCP}
  \alpha_{\ell}\left(\left\| \begin{smallmatrix}x^\diamond- x_s\\ u^\diamond - u_s \end{smallmatrix}\right\|\right) \leq \ell(x^\diamond, u^\diamond) - \ell(x_s, u_s) 
\end{align}
where $(x_s, u_s)\in \mc{S}^{\text{\tiny GNE}}_s$ is the steady-state GNE. By construction, $(x^\diamond, u^\diamond)$ minimizes $\ell(\bar x, \bar u)$. Hence~\eqref{eq:DissAtOCP} holds only if $(x_s, u_s) = (x^\diamond, u^\diamond)$, i.e., the steady-state GNE must coincide with the minimizer of~\eqref{eq:OCPsteadystate}.
This, however, does not hold in general for the GNEP in~\eqref{eq:MPCPerAgent} which asserts our claim.
\end{proof}

\begin{ex}[When do GNEP and OCP coincide?]
We now provide intuition for when the price of anarchy~\eqref{eq:PoA} equals one, focusing on the steady-state problem~\eqref{eq:SteadyStateGNEP} for simplicity.
Indeed, one can identify conditions when $(x_s, u_s) = (x^\diamond, u^\diamond)$ by comparing the first-order optimality conditions of~\eqref{eq:SteadyStateGNEP} and~\eqref{eq:OCPsteadystate};  they coincide if 
\begin{subequations}
\begin{align}
 \frac{\partial\ell^v(x_k,u_k)}{\partial x_k} = \frac{\partial\ell(x_k,u_k)}{\partial x_k}\\[5pt]
 \frac{\partial\ell^v(x_k,u_k^v,u_k^{-v})}{\partial u^{v}_k} = \frac{\partial\ell(x_k,u_k^v,u_k^{-v})}{\partial u^v_k}
\end{align}
\end{subequations}
holds $\forall v\in \mc{V}$ and as $\ell(x_k,u_k) = \sum_{v\in\mc{V}}\ell^v(x_k,u_k)$ this would require for all agents' coupled cost function parts to be the same up to a constant. This would correspond to a potential game~\cite{monderer1996potential} with the sum of costs as the potential function~\cite[Prp.~1]{slade1994what}. We do not impose this restriction in~\eqref{eq:MPCPerAgent}.
\end{ex}

%
%
%

We now formally introduce the considered turnpike property for our GNE problem in~\eqref{eq:MPCPerAgent} with respect to the steady-state GNE~\eqref{eq:SteadyStateGNEP}.
\begin{dfn}[Measure turnpike in GNEPs]\label{dfn:GameTurnpike}~\\
%
The GNEP~\eqref{eq:MPCPerAgent} exhibits the (measure) turnpike property at $(x_s,u_s)$ if for each $\varepsilon>0$ there exists $C>0$ such that $\forall N\in \bb{N}, \forall \x\in  \bb{X}_0$, and for all game pairs $(x,u)\in \mc{S}^{\text{\tiny GNE}}_N(\x)$ it holds that 
\begin{align}\label{eq:TurnpikeInequality}
Q_{\varepsilon} := \#\left\{k \in \bb{Z}_N |\,\left\| \begin{smallmatrix}x_k- x_s\\ u_k - u_s \end{smallmatrix} \right\|\leq \varepsilon\right\} \geq N - \frac{C}{\alpha(\varepsilon)}
\end{align}
for some $\alpha \in \mc K$ and where $\#$ refers to the cardinality.
\end{dfn}
Definition~\ref{dfn:GameTurnpike} provides a lower bound on the number of time steps the GNE trajectory spends within $\mc{B}_\varepsilon(x_s)$, an $\varepsilon$-ball around the steady-state GNE. This bound grows with the horizon $N$, thus intuitively as $N$ increases, the trajectory spends more time in a neighborhood around $x_s$, while the time spent outside $\mc{B}_\varepsilon(x_s)$ remains bounded independently of $N$.


\begin{ass}[Cheap reachability] \label{ass:cheap}
For all $\x\in\bb X_0$, there exists an infinite-horizon feasible pair $(x,u) \in \mc{Z}_\infty(\x)$ 
    such that, for some $\delta \in\R$ and all $N\in \bb N$ it holds that
    \[J_N(x,u)\leq N \ell(x_s, u_s) + \delta.\]
\end{ass}
It is worth noting that this assumption is readily verified if from all $\x\in\bb X_0$, the equilibrium point $x_s$ can be reached in a finite number of steps or if the Jacobian linearization of \eqref{eq:Dynamics} at $(x_s, u_s)$ is stabilizable. Moreover, we note that the feasible $(x,u)$ in Assumption~\ref{ass:cheap} does not need to be a game pair. 

\begin{ass}[Bounded price of anarchy] \label{ass:PoA}~\\
For all $\x\in\bb X_0$, the price of anarchy from \eqref{eq:PoA} satisfies
\[
\text{PoA}(\x) =   \frac{ \displaystyle\sup_{(x^*, u^*) \in \mc{S}^{\text{\tiny GNE}}_N(\x)} J_N(x^*, u^*)}{V^\diamond_N(\mathbf{x})}   \leq P < \infty
\]
for $(x^*,u^*)  \in\mc{S}^{\text{\tiny GNE}}_N(\x)$ and $0 < \nu \leq V^\diamond_N(\mathbf{x}) \leq V< \infty$ holds.
\end{ass}

We next show that strict dissipativity of the GNEP with respect to $(x_s, u_s)$ implies the turnpike property for the input and state trajectories resulting from~\eqref{eq:MPCPerAgent}. The result is inspired by~\cite[Thm. 5.3]{gruene2013economic} and its proof. 

\begin{thm}[Strict dissipativity $\Rightarrow$ turnpike]\label{thm:DissToTurnpike}~\\
Consider the GNEP~\eqref{eq:MPCPerAgent} and let Assumptions~\ref{ass:cheap} and \ref{ass:PoA} hold. Suppose that the GNEP~\eqref{eq:MPCPerAgent} is strictly dissipative with respect to $(x_s,u_s)$ in the sense of Definition~\ref{dfn:StrictDiss} and the storage is bounded on $\bb X_N(\bb X_0)$. Then all GNEP solutions exhibit the (measure) turnpike property at $(x_s,u_s)$.

\end{thm}%
%
%
%
\vspace{-\baselineskip}
\begin{proof}
The proof proceeds in two steps. First we show that Assumptions~\ref{ass:cheap} and \ref{ass:PoA} allows to bound the performance of the population of agents along game pairs. Second we exploit the strict dissipation inequality to construct the asserted bound \eqref{eq:TurnpikeInequality}.

Step 1: \quad Lemma \ref{eq:linearBound} from Appendix~\ref{app:bnd} shows that Assumption~\ref{ass:PoA} implies the linear performance bound
$
J_N(x^*,u^*)   \leq  V^\diamond_N(\x) + \tilde P $ for some finite $\tilde P$.
Since Assumption~\ref{ass:cheap} relies on non-optimal trajectories, 
we have the bound
\[
J_N(x^*,u^*) \leq   N \ell(x_s, u_s) + \delta + \tilde P
\]
which holds for all $N\in \bb N, \x\in\bb X_0$ and $(x^*,u^*)  \in\mc{S}^{\text{\tiny GNE}}_N(\x)$.

Step 2: \quad Let $C:= 2 \sup_{\x\in \bb{X}_0}  |\Lambda(\x)|  < \infty $. Then we have 
\begin{subequations}\label{eq:AveragedShiftedCost}
\begin{align}
\tilde{J}_N(x^*,u^*) &:= J_N(x^*,u^*) + \Lambda(\x) - \Lambda(x_N)\\
&\leq J_N(x^*,u^*) + C, \\ 
&\leq N \,\ell(x_s, u_s) + (\delta + \tilde P  + C)  
\end{align}
\end{subequations}
Also due to strict dissipativity we have that $\, \forall (x^*,u^*) \in \mc{S}^{\text{\tiny GNE}}_N(\mathbf{x}),\, \forall  \mathbf{x}\in \bb{X}_0$: 
\begin{align}\nonumber
\ell(x^*_k,u^*_k) + \Lambda(x^*_k)-\Lambda(f(x^*_k,u^*_k)) \geq \\
 \ell(x_s,u_s) - \alpha_{\ell}\left(\left\|\begin{smallmatrix}x^*_k- x_s\\ u^*_k - u_s \end{smallmatrix}\right\|\right) \ 
\end{align}
Next we construct the bound from \eqref{eq:TurnpikeInequality} with $\alpha = \alpha_\ell$.
For contradiction suppose that the trajectory does not fulfill the turnpike property and thus, $Q_{\varepsilon} < N - \frac{\delta + \tilde C}{\alpha_\ell{(\varepsilon})}$ with $\tilde C = C+\tilde P$. This implies that there exists a number of $N - Q_ {\varepsilon}> \frac{\delta + \tilde C }{\alpha_{\ell}(\varepsilon)}$ times steps in the set $\mc{N}\subseteq \{0, \dots, N-1\}$ for which the trajectory is more than $\varepsilon$ distance away from the steady state $(x_s, u_s)$, i.e., $\|(x^*_k, u^*_k) - (x_s, u_s) \|>\varepsilon, \; \forall k \in \{0,\dots, N-1\}$. We obtain
\begin{align*}
\tilde{J}_N(x^*,u^*) &=  \sum_{k=0}^{N-1} [\ell(x^*_k,u^*_k)) + \Lambda(x^*_k)  - \Lambda(f(x^*_k,u^*_k))]\\
& \geq  N \ell(x_s, u_s) + N (\alpha_{\ell}\left(\left\| \begin{smallmatrix}x^*_k- x_s\\ u^*_k - u_s \end{smallmatrix}\right\|\right) \quad    \text{Def.}\,\ref{dfn:StrictDiss} \\
& = N \ell(x_s, u_s) +  Q_{\varepsilon} \alpha_{\ell}\left(\left\| \begin{smallmatrix}x^*_k- x_s\\ u^*_k - u_s \end{smallmatrix}\right\|\right)]\\
&+ \underbrace{(N - Q_{\varepsilon})\, \alpha_{\ell}\left(\left\|\begin{smallmatrix}x^*_k- x_s\\ u^*_k - u_s \end{smallmatrix}\right\|\right)}_{\geq (N - Q_{\varepsilon} ) \alpha_{\ell}(\varepsilon)} \\
& \geq N \ell(x_s, u_s) + (N - Q_{\varepsilon}) \alpha_{\ell}(\varepsilon) \\
&>  N \ell(x_s, u_s ) + (\delta + \tilde C)
\end{align*} 
which clearly contradicts~\eqref{eq:AveragedShiftedCost}, thus the assertion.
\end{proof}

The proof relies on Assumptions~\ref{ass:cheap} and~\ref{ass:PoA} together with strict dissipativity with bounded storage. Assumption~\ref{ass:PoA}, specific to the GNEP setting, requires that the overall GNEP performance is not arbitrarily worse than its OCP counterpart, which is arguably a mild condition. Assumption~\ref{ass:cheap} is not specific to GNEPs as it is a system property. It can be verified via reachability arguments as done in the OCP turnpike literature~\cite{gruene2013economic}. Similarly, boundedness of the storage function $\Lambda$ is standard in dissipativity-based turnpike analysis. Actually Theorem~\ref{thm:AvailStor} shows that the available storage is always bounded from above and below, so assuming strict dissipativity with bounded storage is no stronger than assuming~\eqref{eq:GsDI} alone. Finally, one might argue that the dissipativity assumption in itself is artificial. Yet, Theorem~\ref{thm:DissToTurnpike} provides an explanation for the turnpike phenomenon in GNEPs and it elegantly links the dynamic game trajectories to the steady-state GNE. Moreover, we will see in Theorem~\ref{thm:TurnpikeToDiss} that also converse statements can be derived: turnpike $\Rightarrow$ dissipativity.


\begin{rmk}[
Turnpike with strict $x$ dissipativity] Observe that---similar to its optimal control counterpart---the proof of Theorem~\ref{thm:DissToTurnpike} does not rely on $(x,u)$-dissipativity with $\alpha_{\ell}\left(\left\| \begin{smallmatrix}x_k- x_s\\ u_k - u_s \end{smallmatrix}\right\|\right) $ but only requires $x$-dissipativity with $ \alpha_{\ell}(\|x_k- x_s\|)$. 
That is, the result of Theorem~\ref{thm:DissToTurnpike} can be slightly generalized to show a measure turnpike at $x_s$ only. 
However, having strictness in $x$ and $u$ will later help to derive converse turnpike results. 
\end{rmk}
\subsection{A converse turnpike result for GNEPs}
We recall the following ``off-equilibrium" concept.
\begin{dfn}[$\rho$-GNE~\cite{chen2021distributed}] \label{def:RhoGNE}
 For a given $\rho \geq 0$, the pair $(u^*, x^*)$ is said to be an $\rho$-generalized Nash equilibrium ($\rho$-GNE) of~\eqref{eq:MPCPerAgent} if $\; \forall v\in \mc{V}$ and  $\forall (x,u^v)\in \mc{Z}_N(\x, u^{-v*})$ it holds:
\begin{align*}
 J_N^v(x^*, u^{v*}, u^{-v*}) \leq J_N^v(x,u^{v}, u^{-v*}) + \rho, \;
\end{align*}
The solution set is denoted as $\mc{S}^{\rho \text{-\tiny GNE}}_N(\x)$. Particularly, when $\rho = 0$, $(u^*, x^*)$ is a GNE and $\mc{S}^{ \rho  \text{-\tiny GNE}}_N(\x)= \mc{S}^{\text{\tiny GNE}}_N(\x)$. 
\end{dfn}

The following assumption ensures that, among all $\rho$-GNEs centred at $x_s$, the pair $(x_s, u_s)$ is the one that minimizes the performance measure $J(x,u)$ defined for the population of agents.

\begin{ass}[Local equilibrium cost bound
]\label{ass:LocalMinimizer}~\\
There exists a constant $\rho>0$ and $\alpha_{\rho} \in \mc{K}$ such that 
$\forall (x,u)\in \mc{S}^{\rho\text{\tiny -GNE}}_N(x_s) \cap \mc{Z}_N(x_s)$ and $\forall k \in \bb{N}$ it holds:
\begin{align}\label{eq:LocalMinimizer}
 \ell(x_s,u_s)   \leq \ell(x_k, u_k)- \alpha_{\rho}\left(\left\| \begin{smallmatrix}x_k- x_s\\ u_k - u_s \end{smallmatrix}\right\|\right) 
\end{align}
where $\mc{Z}_N(x_s)$ is the compact set of all feasible (game) trajectories of length $N$ initialized at $x_s$ as in \eqref{eq:FeasibleSet}.
\end{ass}
In principle, one could also consider $\mc{S}^{\rho_1\text{-\tiny GNE}}_N(\x)$ with a smaller ``suboptimality" gap $\rho_1 \leq \rho$ which still ensures that Assumption~\ref{ass:LocalMinimizer} holds for all trajectories in $\mc{S}^{\rho_1\text{-\tiny GNE}}_N(\x)$.
Assumption~\ref{ass:LocalMinimizer} can be verified by employing the game-theoretic counterpart of Fiacco-like sensitivity analysis. Specifically, in the GNE setting formulated as quasi-variational inequalities, local uniqueness results exist under different sets of assumptions, see~\cite[Ch.~3]{alphonse2021stability} and~\cite[Lem. 7]{noor2025recent}. 
For variational GNEs, sensitivity of variational inequalities can be exploited~\cite[Prop.~12.14]{facchinei2009nash}. 

The following result confirms a converse relation between turnpike and strict dissipativity.
\begin{thm}[Turnpike $\Rightarrow$ strict dissipativity] ~\\Suppose that the stage cost $\ell$ is continuous, that at $(x_s,u_s)$ Assumption~\ref{ass:LocalMinimizer}, and the turnpike property as per Definition~\ref{dfn:GameTurnpike} holds for the GNEP~\eqref{eq:MPCPerAgent}. Then, there exists some $\alpha \in \mc K$, such that the GNEP is strictly dissipative with supply rate $s(x_k,u_k) = \ell(x_k,u_k)-\ell(x_s,u_s)$ with respect to $(x_s,u_s)$. Moreover, there exists a storage function $\Lambda$ bounded on $\bb X_\infty(\bb X_0)$.
\label{thm:TurnpikeToDiss}
\end{thm}
\begin{proof}
Consider $\rho$ for which Assumption~\ref{ass:LocalMinimizer} holds and the corresponding $\alpha_\rho \in \mc{K}$. The turnpike property in~\eqref{eq:TurnpikeInequality} states that given $\rho$\footnote{To improve readability of the proof, we choose $\varepsilon=\rho$ but one may also consider a smaller region around $x_s$ (i.e., $\varepsilon\leq \rho$).} there exists a horizon length $N\in \bb{N}$ for which the GNE trajectory spends nonzero discrete time steps $Q_{\rho}>0$ within $\mc{B}_\varepsilon(x_s)$. Further, recall the definition of the available storage from Theorem~\ref{thm:AvailStor}
%
\begin{align}\label{eq:AvailStor}
{\Lambda}_{\alpha_\rho}(\x) =  \sum_{k=0}^{N-1} \alpha_\rho\left(\left\|\begin{smallmatrix}x_k^*- x_s\\ u_k^* - u_s \end{smallmatrix}\right\|\right) - s(x_k^*,u_k^*)
\end{align}
evaluated along the game pair $(x^*, u^*)\in  \mc{S}^{\text{\tiny GNE}}_N(\x)$. Here, without loss of generality, we use the $\mc{K}$-class function $\alpha_{\rho}$ from Assumption~\ref{ass:LocalMinimizer}. 

Temporarily, consider some $N \in \mathbb N$ and a feasible initial condition $\x$. We split the time horizon into $Q_{\rho}(N,\x)$, the trajectory points close to $(x_s,u_s)$, and 
\[Q_{out}(N,\x):= \{0,\dots,N-1\}\backslash Q_\rho(N,\x),\] 
i.e. the points outside of $\mc{B}_\rho(x_s)$. Evaluating~\eqref{eq:AvailStor} on $Q_{out}(N,\x)$ gives
%
\begin{align}\label{eq:StorageOut} \nonumber
    &\sum_{k\in Q_{out}(N,\x)} \alpha_\rho\left(\left\| \begin{smallmatrix}x_k^*- x_s\\ u_k^* - u_s \end{smallmatrix}\right\|\right) - s(x_k^*,u_k^*) \\ 
    &\leq \#Q_{out}(N,\x) \cdot (\hat \alpha + \hat{s})
\end{align}
where \[\displaystyle \hat{\alpha} = \sup_{(x,u)\in \mc{Z}_N(\x)} \alpha_\rho\left(\left\|\begin{smallmatrix}x_k- x_s\\ u_k - u_s \end{smallmatrix}\right\|\right)\] and similarly $ \displaystyle \hat{s} = \sup_{(x,u)\in \mc{Z}_N(\x)}  s(x_k,u_k)$. The turnpike property implies that any trajectory, independent of $N$ and $\x$, only spends a finite amount of time outside of the ball $\mc{B}_\rho(x_s)$. This allows us to establish the bound
\begin{align}
     \#Q_{out}(N,\x) \cdot (\hat \alpha + \hat{s})\leq \bar Q_{out}\cdot (\hat \alpha + \hat{s})<\infty
\end{align}
where due to \eqref{eq:TurnpikeInequality} $\bar Q_{out} \geq \frac{C}{\alpha_\rho(\rho)}$ is some constant independent of $N$ and $\x$. 

Next, we evaluate the second part of the trajectory. For $Q_{\rho}(N,\x)$ we have that due to the turnpike property $\forall k\in Q_{\rho}(N,\x)$  it holds that $\|x_k-x_s\|\leq \rho$. Assumption~\ref{ass:LocalMinimizer} holds $\forall k\in Q_{\rho}(N,\x)$ by choice of $\rho$. Thus we can construct the following inequality:
\begin{align}\label{eq:StorageIn}
    \sum_{k\in Q_{\rho}(N,\x)} \alpha_\ell\left(\left\| \begin{smallmatrix}x_k^*- x_s\\ u_k^* - u_s \end{smallmatrix}\right\|\right) - s(x_k^*,u_k^*)\leq 0
\end{align}
where we use the fact that $s(x_k,u_k) = \ell(x_k,u_k) - \ell(x_s,u_s)$ .
Combining~\eqref{eq:StorageOut} and~\eqref{eq:StorageIn} we get that 
\begin{align}\nonumber
&\sum_{k=0}^{N-1} \alpha_\rho\left(\left\| \begin{smallmatrix}x_k^*- x_s\\ u_k^* - u_s \end{smallmatrix}\right\|\right)- s(x_k^*,u_k^*) \\
&\leq \bar Q_{out} \cdot (\hat \alpha + \hat{s})<\infty \label{eq:AvailStorBound}
\end{align}
Thus, as $ \bar Q_{out} \cdot (\hat \alpha + \hat{s})$ is independent of $N$, \eqref{eq:AvailStorBound} holds for any $\x$ for which $(x^*, u^*) \in \mc{S}^{\text{\tiny GNE}}_N(\x)$ and constitutes a bound on the available storage. As per Theorem~\ref{thm:AvailStor}, GNEP~\eqref{eq:MPCPerAgent} is strictly dissipative with respect to $(x_s,u_s)$.

Lastly, the existence of a bounded storage function follows from the observation that the available storage is bounded on $\bb X_\infty(\bb X_0)$, cf. Appendix~\ref{apendix:ProofThmAvailStor}.
\end{proof}

This theorem highlights the close relation of dissipativity and turnpike properties in GNEPs. Theorems~\ref{thm:DissToTurnpike} and \ref{thm:TurnpikeToDiss} show that under rather mild conditions, strict GNEP dissipativity in the sense of Definition~\ref{dfn:StrictDiss} and the turnpike property from Definition~\ref{dfn:GameTurnpike} are equivalent. 

\begin{cor}[Turnpike $\Leftrightarrow$ strict dissipativity]
   Consider the GNEP~\eqref{eq:MPCPerAgent} and let Assumptions~\ref{ass:cheap}--\ref{ass:LocalMinimizer} hold. Suppose that the stage cost $\ell$ is continuous and let $\bb X_N(\bb X_0) = \bb X_0$. Then the following two statements are equivalent:
   \begin{enumerate}
         \item[(i)] The GNEP~\eqref{eq:MPCPerAgent} is strictly dissipative with bounded storage (Definition~\ref{dfn:StrictDiss}). 
       \item[(ii)] The GNEP~\eqref{eq:MPCPerAgent} has the measure turnpike property (Definition~\ref{dfn:GameTurnpike}).
    \end{enumerate}
\end{cor}
Due to Theorems~\ref{thm:DissToTurnpike} and \ref{thm:TurnpikeToDiss}, the proof is straightforward and thus omitted. The condition $\bb X_N(\bb X_0) = \bb X_0$ is imposed to match the set of initial conditions for which Assumption~\ref{ass:cheap} holds with the domain of the available storage. This condition is satisfied if from any point on a GNEP solution trajectory the turnpike $(x_s, u_s)$ can be reached in finitely many steps.


\subsection{Optimal game operation at steady-state GNEs}

We have established in the previous section under which conditions the open-loop trajectory will spend most of its time at the turnpike, the steady-state GNE. However, this raises the question if the pair $(x_s,u_s)$ is also the best performing GNE (of potentially very many GNEs) in terms of the entire population performance measure $J_N(x,u)$ from \eqref{eq:overallJ}. To investigate this, we define optimal steady state operation and suboptimal operation, following~\cite[Def. 2]{mueller2015role}.
\begin{dfn}[Optimal game operation]~\\
The GNEP~\eqref{eq:MPCPerAgent} is \textit{optimally operated at a steady-state GNE}, if for each $\mathbf{x}\in \bb{X}_0$ and each game pair $(x,u)\in \mc{S}^{\text{\tiny GNE}}_N(\mathbf{x}) $ the following holds $\forall k\in \mathbb{I}_{\geq 0 }$
\begin{align}\label{eq:OptOperation}
\liminf_{N\to +\infty} \frac{\sum_{k=0}^{N-1}\ell(x_k,u_k)}{N} \geq \ell(x_s,u_s)
\end{align}
for $(x_s,u_s) \in \mc{S}_s^{\text{\tiny GNE}}(\mathbf{x})$.
It is \textit{suboptimally operated off steady state GNE} if~\eqref{eq:OptOperation} holds in a strict sense.
\end{dfn}
Optimal steady-state GNE operation implies that no feasible game pair of~\eqref{eq:MPCPerAgent} can have a better (asymptotic) average performance than the performance of the steady-state GNE $(x_s, u_s)$, and suboptimal operation off steady-state means the performance is strictly worse or ``passes by'' $x_s$ infinitely often. Next, we relate strict dissipativity of~\eqref{eq:MPCPerAgent} to optimal operation at $(x_s,u_s)$.
\begin{prp}\label{prop:OptimalOperation}
Suppose that the set $\mc{S}_{\infty}^{GNE}(\x) \not= \emptyset $, then if GNEP\,\eqref{eq:MPCPerAgent} is strictly dissipative with respect to $(x_s,u_s)$ following Definition~\ref{dfn:StrictDiss} and with bounded storage, it is optimally operated at the steady-state GNE $(x_s,u_s)$ and suboptimally operated off steady state.
\end{prp}

\begin{proof} The proof follows along the lines of~[Prop. 6.4]\cite{angeli2012average}. First, we note that $\displaystyle \sup_{x\in \bb{X}_0}  |\Lambda(x)| < \infty $ giving a bound on the storage function and allowing us to state the following equality 
\begin{align*}
0 &= \lim_{N\to +\infty} \frac{\Lambda(x_N) - \Lambda(x_0)}{N}\\
&= \lim_{N\to +\infty} \frac{\sum_{k=0}^{N-1} \Lambda(x_{k+1}) -\Lambda(x_k)}{N}\\
&\leq  \liminf_{N\to +\infty} \frac{ \sum_{k=0}^{N-1}\ell(x_k,u_k) - \alpha_{\ell}\left(\left\| \begin{smallmatrix}x_k- x_s\\ u_k - u_s \end{smallmatrix}\right\|\right)}{N} - \ell(x_s,u_s)
\end{align*}
Note that switching to $\liminf$ is necessary as the sequence $\ell(x_k,u_k)$ may not be converging.
Using superadditivity\footnote{Superadditivity: $ \liminf\limits_{n\to \infty}(a_n + b_n)\geq \liminf\limits_{n\to \infty} a_n +  \liminf\limits_{n\to \infty}b_n$} of the $\liminf$ and noting that $\ell(x_s, u_s)$ is constant we obtain
\begin{align*}
&\liminf_{N\to +\infty} \frac{\sum_{k=0}^{N-1}\ell(x_k,u_k)}{N} \\
&\geq \liminf_{N\to +\infty} \frac{ \sum_{k=0}^{N-1}\alpha_{\ell}\left(\|\begin{smallmatrix}x_k- x_s\\ u_k - u_s \end{smallmatrix}\|\right)}{N} + \ell(x_s,u_s).
\end{align*}
We consider an infinite horizon feasible game pair  $(x^*,u^*) \in  \mc{S}^{\text{\tiny GNE}}_{\infty}(\x)$ and distinguish two cases: \\
(i) If $x_k \not \to x_s$ as $N\to \infty$, then 
\[ \liminf_{N\to +\infty} \frac 1 N\left(\sum_{k=0}^{N-1}\ell(x_k,u_k)\right) > \ell(x_s,u_s)\]
due to the positive definiteness $\alpha_{\ell}(\cdot)$ proving suboptimal operation off steady state;\\
(ii) Else if  $\displaystyle  \liminf_{N\to +\infty} \|x_k- x_s\|=0$ and $\displaystyle  \liminf_{N\to +\infty} \|u_k- u_s\|=0$ then
\[\displaystyle \liminf_{N\to +\infty}  (\ell(x_k,u_k))/N) = \ell(x_s,u_s)\] proving optimal operation at steady state. \end{proof}

To the best of our knowledge, Proposition 1 is the first characterization of GNEs from an optimality perspective. It shows that if the group of agents converges to $(x_s,u_s)$ (i.e., strict dissipativity holds), then this point is the best possible (hence optimal) steady-state GNE for the population in terms of cost. \footnote{Note that such a characterization exists for potential games \cite{monderer1996potential} that have a direct connection to optimization problems through the potential function. Yet, in~\eqref{eq:MPCPerAgent} we consider a general class of GNEPs.}

\begin{figure}
        \centering
        \includegraphics[width=\linewidth]{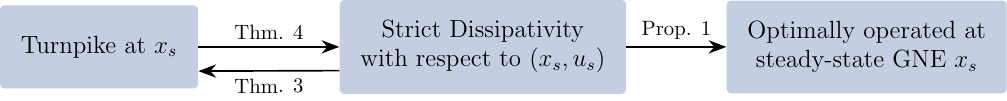}
        \caption{Overview of implications between strict dissipativity, turnpike, and optimal
operation at the steady-state GNE.}
  \end{figure}  
%
\section{Value Function and KKT Interpretation}\label{sec:OCP-GNE-relation}
In this section we aim to draw a connection between the GNE solution of~\eqref{eq:MPCPerAgent} and their KKT conditions. In order to do so, we state the following assumptions on the functions in the GNEP~\eqref{eq:MPCPerAgent}.

\begin{ass}[Differentiable problem data] \label{ass:ContDiff}
For every $v\in\mc{V}$ and fixed $u^{-v}$, the cost $J_N^v(\cdot,\cdot, u^{-v})$, the constraint functions $g(\cdot, \cdot, u^{-v}), h^v(\cdot),$ and the dynamics $f(\cdot,\cdot, u^{-v})$ are continuously differentiable in $x$ and $u^v$. 
\end{ass}

\begin{ass}[Convexity] \label{ass:Convex}For every $v\in\mc{V}$ and fixed $u^{-v}$, the cost  $J^v(\cdot,\cdot, u^{-v})$
is convex and the set $\mc{Z}_N^v( \cdot, u^{-v})$ is closed and convex.
\end{ass}

The convexity assumption is required to ensure that points which solve the individual agent's KKT system are not merely stationary points (local minima, saddle points). KKT conditions of GNEs and relevant constraint qualifications have been studied in detail in~\cite{bueno2019optimality}.

At each time step $k = 0,\ldots, N-1$ consider
 \[
 L^v_k = \begin{bmatrix}
     1 \\\lambda^v_{k+1}\\\mu_{k}^v \\ \nu_{k}^v      
 \end{bmatrix}^\top
 \begin{bmatrix}
 \ell^v(x_k,u_k, u^{-v}_k) \\  f(x_{k},u_{k}^v, u_k^{-v}) - x_{k+1} \\
 g(x_k,u_k^{v},u^{-v}_k) \\ h^v(u_k^v) 
  \end{bmatrix}.
 \]
The Lagrangian of each agent associated with~\eqref{eq:MPCPerAgent} can then be written as
\begin{multline}\label{eq:Lagrangian}
\mc L^v(x,u, \lambda^v, \mu^v, \nu^v) = \sum_{k=0}^{N-1} L_k^v
 +(\lambda_{0}^v)^\top [\x - x_0]
 \end{multline}
This yields the following per-agent KKT system which needs to hold $\forall v\in \mc{V}$ and $k\in \bb{N}$
\begin{subequations} \label{eq:KKT_GNEP}
\begin{align} 
  x_{k+1} &= f(x_k,u_k^v, u_k^{-v}) \label{eq:KKT_GNEPx}\\
%
%
\lambda^v_{k} &= \ell_x^v +  g_{x}^\top \mu^v_{k} + f_x^\top \lambda_{k+1}^v, \quad \\ 
%
%
0 &=   \ell_{u^v}^v  +  g_{u^v}^\top \mu^v_{k}  + f_{u^v}^\top \lambda_{k+1}^v +  h_{u^v}^\top \eta_k^v, \label{eq:KKT_GNEPu}
\end{align} 
subject to the initial and boundary conditions 
\begin{equation}\label{eq:KKT_GNEPbnd}
    x_0 = \x \quad \text{and}\quad \lambda^v_{N} = g_{x}^\top \mu^v_{N}
\end{equation}
and the usual conditions of primal feasibility and complementarity slackness
\begin{align}\label{eq:KKT_GNEPslack1}
0 \leq \mu_k^v &\quad \perp \quad  -g^v(x_k,u_k^v,u_k^{-v})\geq 0,\\
0 \leq \eta_k^v&\quad\perp\quad -h^v(u^v_k) \geq 0.\label{eq:KKT_GNEPslack2}
\end{align}
\end{subequations}

%
%
%
%

A well-established result from the GNE literature connects this KKT system to GNEs, cf.~\cite[Thm. 4.6]{facchinei2009generalized}.
\begin{thm}[KKT conditions for GNEPs]\label{thm:KKTGNEP}
Given Assumption~\ref{ass:ContDiff} then the following statements hold for the GNEP~\eqref{eq:MPCPerAgent}:
\begin{enumerate}
\item[(i)] Let $(x^*,u^*)$ be an equilibrium of the GNEP at which all agents' subproblems satisfy Slater's constraint qualification. Then, $\forall v\in \mc{V}: \exists\,  (\lambda^{v*},\mu^{v*},\eta^{v*})$ that together with $(x^*,u^*)$  solve the KKT system~\eqref{eq:KKT_GNEP}.
\item[(ii)] If $(x^*,u^*)$ and $ (\lambda^{v*},\mu^{v*},\eta^{v*})_{v\in\mc{V}}$ solve the KKT system~\eqref{eq:KKT_GNEP} and Ass.~\ref{ass:Convex} holds, then $(x^*,u^*)$ is an equilibrium point of the GNEP. 
\end{enumerate}
\end{thm}

Similar to~\cite{faulwasser2018asymptotic}, we study when a primal-dual solution to the steady-state GNEP~\eqref{eq:SteadyStateGNEP} is also a solution for the dynamic counterpart~\eqref{eq:MPCPerAgent}. To this end, we consider the Lagrangian for each agent of the steady-state GNE problem in~\eqref{eq:SteadyStateGNEP} given by
\begin{align*}
&\mc L_s^v(\bar{x},\bar{u}, \bar{\lambda}^v, \bar{\mu}^v, \bar\nu^v) =  \ell^v(\bar{x},\bar{u}^v, \bar u^{-v}) + (\bar{\mu}^v)^\top\,  g(\bar{x},\bar{u}^v, \bar u^{-v}) \\
& + (\bar{\lambda}^v)^\top[ f(\bar{x},\bar{u}^v, \bar u^{-v}) -\bar{x}] + (\bar \eta^v)^\top h^v(\bar u^v).
\end{align*}
Computing the usual stationarity condition with respect to $(\bar{x},\bar{u}, \bar{\lambda}^v, \bar{\mu}^v, \bar{\nu}^v)$ gives  the following set of  conditions
\begin{subequations}\label{eq:steadystateGNEKKT}
\begin{align} 
0 &= f(\bar{x},\bar{u}^v, \bar u^{-v}) - \bar{x}, \label{eq:steadystateGNEKKT_x} \\
%
  \bar{\lambda}^v &= \ell_x^v +  g_{x}^\top \bar{\mu}^v + f_x^\top \bar{\lambda}^v, \label{eq:steadystateGNEKKT_lbd}\\
%
0 &=   \ell_{u^v}^v  +  g_{u^v}^\top \bar{\mu}^v + f_{u^v}^\top \bar{\lambda}^v +  h_{u^v}^\top  \bar{\eta}^v , \label{eq:steadystateGNEKKT_u}
%
\end{align} 
which together with 
\begin{align}
 0 \leq \bar\mu^v &\quad\perp\quad  -g^v(\bar x,\bar u^v, \bar u^{-v}) \geq 0,\\
 0 \leq \bar\eta^v &\quad\perp\quad -h^v(\bar u^v) \geq 0,
\end{align}
\end{subequations}
form the KKT conditions for~\eqref{eq:SteadyStateGNEP}. 

\subsection{The game value function of GNEPs}
In the analysis of optimal control problems, the value function plays a fundamental role. It allows to characterize optimality in Dynamic Programming~\cite{Bellman54a}, it can also be linked to the Maximum Principle. Specifically, in first-order optimality conditions of the Maximum Principle the gradient of the OCP value function is related to the co-state/adjoint, i.e., to the dual variable linked to the equality constraints imposed by the dynamics.  Given that~\eqref{eq:KKT_GNEP} are from a per-agent perspective just first-order (KKT) conditions---though all agents consider the same dynamics and state trajectory---one may wonder is there a GNEP-specific counterpart of the optimal control value function? 


To this end, recall the construction of $\ell$ as the sum over all agent-specific costs $\ell^v$, cf.~\eqref{eq:ellStack}. Summing $\ell$ from $k=0$ to $k=N-1$ leads to the 
\textit{game value function} $V_N^*:\bb X_0 \to \R$ which measures the performance of the agent population at $\x$
\begin{equation}\label{eq:Vfun}
  \hspace{-1.5mm}  V^*_N(\x) := \sum_{k=0}^{N-1}\ell(x_k^*, u^*_k)
    = \sum_{v\in \mc V}\sum_{k=0}^{N-1}\ell^v(x_k, u_k^v, u_k^{-v}).
\end{equation}

\begin{thm}[Sensitivity characterization]\label{thm:Vgrad}
    Consider the GNEP~\eqref{eq:MPCPerAgent}, where all inequality constraints are neglected, and let Assumptions~\ref{ass:ContDiff}--\ref{ass:Convex} hold.
    Suppose that for some $\varepsilon >0$ and for any perturbed initial condition
    \[
    x_0 = \x +\xi, \quad \xi \in \mc B_\varepsilon(\x)
    \]
    we have $\mc S^{GNE}_N(\x+\xi)\not= \emptyset$, i.e., the GNEP~\eqref{eq:MPCPerAgent} admits a solution such that $V^*_N(\x)$ is locally differentiable at $\x$. 
    
    Then 
    \begin{equation}\label{eq:gradVfun}
        \nabla V^*_N(\x) = \sum_{v\in\mc V} \lambda^v_0,
    \end{equation}
    where $\lambda^v_0$ corresponds to the KKT solution tuple $(x, u, \lambda^v, 0, 0)$ with $(x, u)\in\mc S^{GNE}_N(\x)$.
\end{thm}
\begin{proof}

Consider the GNEP~\eqref{eq:KKT_GNEP} without inequality constraints and subject to the perturbed initial condition
$
x_0 = \x +\xi, \quad \xi \in \mc B_\varepsilon(\x) \subset \mathbb R^n$.
For all $v\in \mc V$, let $(x(\xi), u^v(\xi), \lambda^v(\xi))$ denote the solution to the KKT conditions~\eqref{eq:KKT_GNEP} with perturbed initial condition, which for all $\xi \in \mc B_\varepsilon(\x)$ are dynamic GNEs due to Theorem~\ref{thm:KKTGNEP}. 

The solution for the unperturbed initial condition is 
\[
(x^*, u^{v,*}, \lambda^v) = (x(0), u^v(0), \lambda^v(0)).
\]
Computing the sensitivity of the Lagrangian~\eqref{eq:Lagrangian} with respect to $\xi$ at $(x(\xi), u^v(\xi), \lambda^v(\xi))$
gives
\begin{multline*}
\dfrac{\mathrm d }{\mathrm d \xi }\mc L^v\left(x(\xi), u^v(\xi), \lambda^v(\xi)\right) =\\
\sum_{k=0}^{N-1}\dfrac{\partial  L_k^v}{\partial x_k }\dfrac{\partial x_k}{\partial\xi } + \dfrac{\partial  L_k^v}{\partial u^v_k }\dfrac{\partial u^v_k}{\partial\xi } + \dfrac{\partial  L_k^v}{\partial \lambda^v_k }\dfrac{\partial \lambda^v_k}{\partial\xi }+\lambda_0^v(\xi)^\top  \dfrac{\partial x_0}{\partial\xi },
\end{multline*}
where $\frac{\partial x_k}{\partial\xi }$ is the sensitivity of state $x_k(\xi)$ to the perturbation of the initial condition $x_0 = \x +\xi$. Likewise  $\frac{\partial u^v_k}{\partial\xi }$ and $\frac{\partial \lambda^v_k}{\partial\xi }$ are sensitivities with respect to $\xi$. 

Notice that in the KKT conditions~\eqref{eq:KKT_GNEP} the optimality-like system~\eqref{eq:KKT_GNEPx}--\eqref{eq:KKT_GNEPu} follows from  \[\nabla_{x_k}L^v_k=0, \nabla_{u^v_k}L^v_k=0, \quad \text{and} \quad\nabla_{\lambda^v_k}L^v_k=0.\]
Hence, for all $v\in \mc V$, we obtain
\begin{align}\label{eq:1stVariation}
   \hspace{-0mm}\left.\dfrac{\mathrm d }{\mathrm d \xi }\mc L^v(x(\xi), u^v(\xi), \lambda^v(\xi))\right|_{\xi = 0} &= 
\left.\lambda_0^v(\xi)^\top  \dfrac{\partial x_0}{\partial\xi }\right|_{\xi = 0} \nonumber\\&= \lambda^v(0)^\top.
\end{align}
Observe that the definition of $V^*_N(\x)$ in~\eqref{eq:Vfun} implies
\[
V^*_N(\x+\xi) = \sum_{v\in \mc V}\mc L^v(x(\xi), u^v(\xi), \lambda^v(\xi)).
\]
differentiating on both sides with respect to $\xi$ gives
\[
\dfrac{\mathrm d}{\mathrm d \xi }  V^*_N(\x+\xi)=  \sum_{v\in \mc V}\dfrac{\mathrm d }{\mathrm d \xi }\mc L^v(x(\xi), u^v(\xi), \lambda^v(\xi)).
\]
Setting $\xi=0$ and using~\eqref{eq:1stVariation} proves the assertion. 
\end{proof}

Sensitivity characterization of the value function is an important step towards a more general characterization of value functions in GNEPs with self-interested agents and allows to draw connections to the storage function.

\begin{figure*}[t]
    \centering
    \includegraphics[width=1\textwidth]{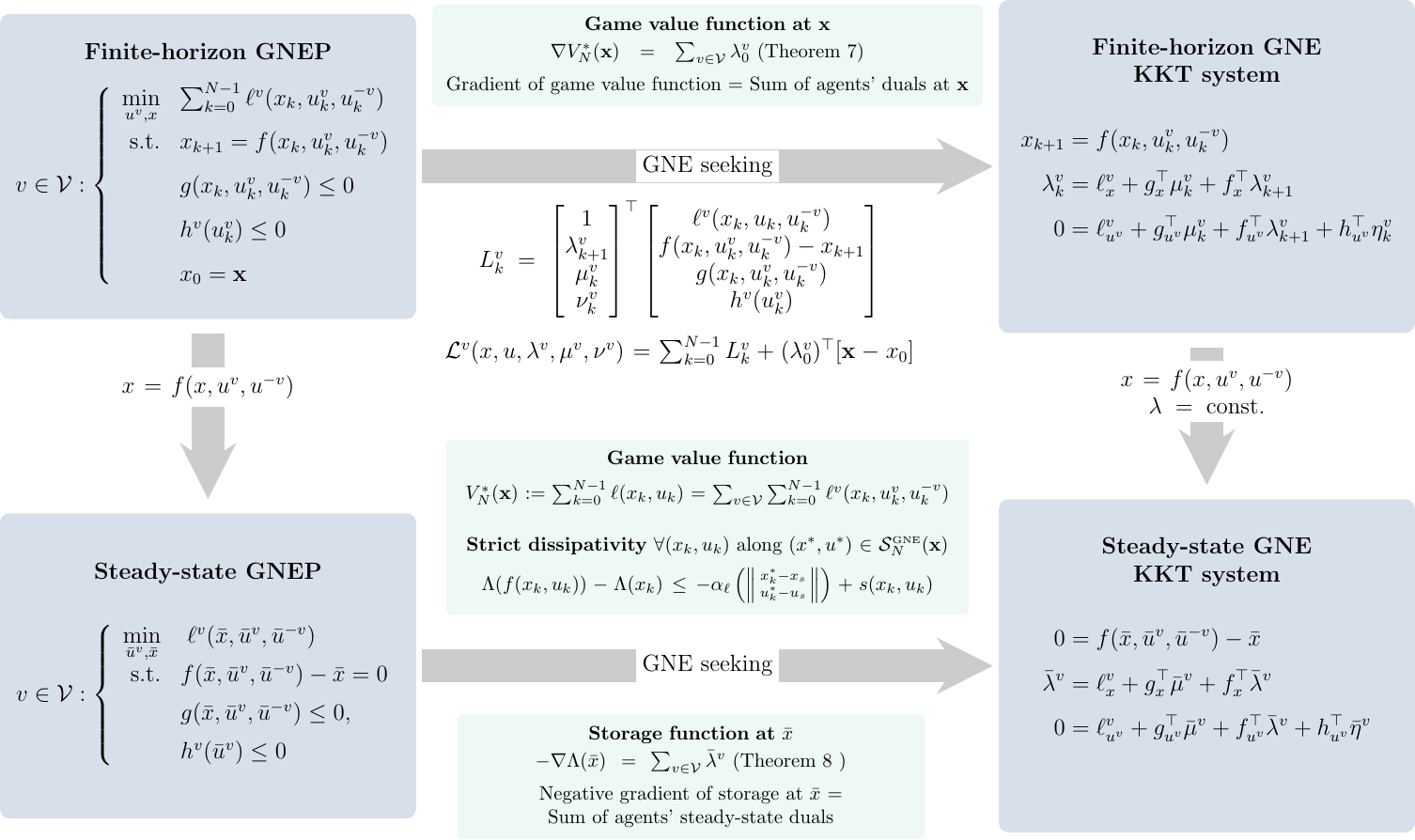}
    \caption{Finite-horizon and steady-state GNEP with GNE-KKT system}
    \label{fig:gnep-kkt}
\end{figure*}

\subsection{Sensitivity analysis of the storage function}
The next result translates a crucial observation made in the context of turnpike analysis of OCPs 
to GNEPs providing a storage function counterpart to Theorem~\ref{thm:Vgrad}.

%

\begin{thm}[Storage gradient]\label{thm:storGrad}
 Consider GNEP~\eqref{eq:MPCPerAgent}, its steady-state counterpart~\eqref{eq:SteadyStateGNEP}, and let Assumptions~\ref{ass:ContDiff}--\ref{ass:Convex}  hold. Then, the following statements hold:
 \begin{enumerate}[(i)]
     \item Any primal-dual solution $(\bar{x},\bar{u}, \bar{\lambda}^v, \bar{\mu}^v, \bar\nu^v)$ of~\eqref{eq:steadystateGNEKKT} constitutes a steady-state solution of  (\ref{eq:KKT_GNEPx}--\ref{eq:KKT_GNEPu}).
     \item Let  $(\bar{x},\bar{u}, \bar{\lambda}^v, \bar{\mu}^v=0, \bar{\nu}^v)$   solve~\eqref{eq:steadystateGNEKKT} and $\bar x \in\bb X_N(\bb X_0)$ holds.
     Then, for any storage function $\Lambda$ differentiable at $\bar x$, it holds that
       \begin{equation}\label{eq:storGrad}
       \sum_{v\in \mc V}\bar\lambda^v = - \nabla \Lambda(\bar x).
     \end{equation}
 \end{enumerate}
\end{thm}
\begin{proof}
    \textit{Part (i):} Notice that at steady state we can neglect the initial condition for $x$ and the terminal condition for $\lambda^v$. Then, assertion (i) follows from direct comparison of~\eqref{eq:steadystateGNEKKT} and~\eqref{eq:KKT_GNEP}. 
   
   \textit{Part (ii):}
%
 %
    The first sentence of part (ii) gives that the strict dissipation inequality~\eqref{eq:GsDI} can be evaluated at $(\bar{x},\bar{u})$.
   Indeed, at $(\bar{x},\bar{u})$, the strict dissipation inequality~\eqref{eq:GsDI} holds with equality. Differentiating it at $\bar x$ gives
    \begin{equation}\label{eq:nabla_sDI}
       0 = \ell_x -  \left(f_x^\top - I\right)\nabla\Lambda(\bar x)
    \end{equation}
    with $\ell$ from~\eqref{eq:ellStack}.
    Now sum~\eqref{eq:steadystateGNEKKT_lbd} over $v \in \mc V$ to obtain
    \begin{equation*}\label{eq:sum_KKTGNE_lbd}
    0 = \sum_{v\in \mc V}\ell_x^v+\left(f_x^\top - I\right)\lambda^v
    = \ell_x + \left(f_x^\top - I\right)\sum_{v\in \mc V}\bar\lambda^v
    \end{equation*}
    where we use~\eqref{eq:ellStack} and the fact that $\bar{\mu}^v=0$.
    Comparison with~\eqref{eq:nabla_sDI} directly gives the assertion.       
\end{proof}

Note that Theorem~\ref{thm:storGrad}~(i) states that (\ref{eq:steadystateGNEKKT_x}--\ref{eq:steadystateGNEKKT_u}) is a fixed point for (\ref{eq:KKT_GNEPx}--\ref{eq:KKT_GNEPu}) but not for~\eqref{eq:MPCPerAgent} as this would also require the boundary conditions~\eqref{eq:KKT_GNEPbnd} to hold. 
Moreover, the negative sign in~\eqref{eq:storGrad} stems from the steady-state constraint $f(\bar{x},\bar{u}^v,\bar{u}^{-v})- \bar{x}=0$ in~\eqref{eq:SteadyStateGNEP}. If this constraint is equivalently written as $\bar{x} -f(\bar{x},\bar{u}^v,\bar{u}^{-v})=0$, then the corresponding Lagrange multiplier flips sign and the minus in~\eqref{eq:storGrad} can be dropped. 

 If $\bar{\mu}^v\not=0$, there exist active per-agent constraints in the steady-state problem~\eqref{eq:SteadyStateGNEP}. The previous result can then be extended to 
\[
       \sum_{v\in \mc V}\bar\lambda^v + \left(f_x^\top - I\right)^\dagger g_x^\top\bar\mu^v= - \nabla \Lambda(\bar x),
     \]
where $(\cdot)^\dagger$ denotes the Moore-Penrose inverse. 

The above result shows two important links between the GNEP~\eqref{eq:MPCPerAgent} and its steady-state counterpart~\eqref{eq:SteadyStateGNEP}: The optimality systems of both problems are closely related which resembles insights from optimal control theory~\cite{Trelat15a,zanon2018economic,faulwasser2022turnpike}. Moreover, there exists a strong link between the dual variables $\lambda^v$ and the storage function $\Lambda$ which has been previously shown for OCPs~\cite{Stieler14a,zanon2018economic,faulwasser2018asymptotic}. Yet, to the best of our knowledge in the context of GNEPs this link is novel. 

 We conclude this section with a corollary that evaluates~\eqref{eq:storGrad} at the turnpike state $x_s$.
 \begin{cor}
     Consider the setting of Theorem~\ref{thm:storGrad}. If $\bar x=x_s$ we have that
     \[ \sum_{v\in \mc V}\lambda_s^v = - \nabla \Lambda(x_s),\]
     where, for all $v\in \mc V$, $(x_s, u_s^v, \lambda_s^v, \mu_s^v, \nu_s^v)$ is a KKT point solving~\eqref{eq:steadystateGNEKKT}.
 \end{cor}

If the GNEP exhibits a turnpike at $(x_s, u_s)$ and the horizon $N$ is sufficiently large, we arrive at a crucial link between the value function $\eqref{eq:Vfun}$ and any differentiable storage function
\[
\nabla V^*_N(x_s) = \sum_{v\in\mc V} \lambda^v_0  \approx \sum_{v\in \mc V}\lambda_s^v = - \nabla \Lambda(x_s)
\]
or, for all $v\in \mc V$, $\lambda^v_0 \approx \lambda_s^v$. The horizon being sufficiently large ensures that the GNEP trajectory converges sufficiently close to $x_s$. 

\subsection{Turnpike leaving arc and linear end penalties}\label{subsec:LinearEndPenalty}

A well-known aspect of the turnpike property is the characteristic leaving arc, where state and input trajectories diverge from the turnpike in the final steps of the horizon, cf. Figure~\ref{fig:Turnpike_Schematic}. While the Definition~\ref{dfn:GameTurnpike} does not require a leaving arc, the early  works~\cite{Dorfman58,Mckenzie76} and later results~\cite{Anderson87a,Trelat15a} focus on it. 

In the context of optimal control, the presence of a leaving arc arises when the values of the states $x$ and the adjoints $\lambda$ close to the turnpike differs substantially from their values at the end of the horizon. In optimal control there are three mechanisms which generate leaving arcs~\cite{faulwasser2022turnpike}:
\begin{itemize}
    \item Terminal constraints on the state variable such that $x_N = x^N \not = x_s$. Naturally this can be generalized to $x_N\in \mathbb X_N$.
    \item Terminal stage costs $V_\mathrm{f}(x)$ which (without any active constraints at $k=N$) imply the transversality condition $\lambda_N = \nabla V_\mathrm{f}(x)$.
    \item Or the presence of state constraints and a cost which renders it cheap to let a system drift towards the constraint boundary at the end of the horizon. 
\end{itemize}

Interestingly, the same leaving arc phenomenon can be observed in GNEPs, see, e.g., the supply chain example~\cite{hall2024receding}. Moreover, the optimality interpretation of GNEPs---and in particular the structure of the KKT conditions, which is very similar to classic optimal control---suggest that the same mechanisms generate a leaving arc. 

We have established in Proposition~\ref{prop:OptimalOperation} that under a strict dissipativity assumption the steady-state GNE in~\eqref{eq:SteadyStateGNEP}. Thus, it is desirable to enforce that agents not only converge to $(x_s,u_s)$ but remain there.

Inspired by the OCP analysis in~\cite{faulwasser2018asymptotic}, the next result shows how to suppress leaving arcs in GNE trajectories. 

\begin{prop}[GNEPs without leaving arcs]\label{prop:NoLeavArc}
Consider the GNEP~\eqref{eq:MPCPerAgent} and let Assumptions~\ref{ass:ContDiff}--\ref{ass:Convex} hold. 
Let $(x_s, u_s^v, \lambda_s^v, \mu_s^v, \nu_s^v)$ be a KKT point solving~\eqref{eq:steadystateGNEKKT}.
Let one of the following two statements hold:
\begin{enumerate}[(i)]
    \item For all $v\in \mc V$, in~\eqref{eq:MPCPerAgent} the per-agent cost function is of the form 
\[
\sum_{k= 0}^{N-1} \ell^v(x_k, u_k^v,u_k^{-v}) + V_\mathrm{f}^v(x_N)
\]
and  
$\nabla V_\mathrm{f}^v(x_s) + g_x^\top\mu_N^v = \lambda_s^v$.
\item For all $v\in \mc V$, in~\eqref{eq:MPCPerAgent} the point-wise terminal constraint $x_N = x_s$ is considered.
\end{enumerate}
Then, for all $N\in \mathbb N_{>0}$, the constant trajectory $(x, u) \equiv (x_s, u_s)$ satisfies
\[(x_s, u_s)\equiv: (x, u) = (x^*, u^*) \in \mc S_N^{GNE}(x_s). \] 
\end{prop}
\begin{proof}
As per Theorem~\ref{thm:storGrad}, part (i) the steady-state primal-dual tuple $(x_s, u_s^v, \lambda_s^v, \mu_s^v, \nu_s^v)$ from~\eqref{eq:steadystateGNEKKT} is a steady state of (\ref{eq:KKT_GNEPx}--\ref{eq:KKT_GNEPu}). Moreover, using $(x_s, u_s^v, \lambda_s^v, \mu_s^v, \nu_s^v)$ to construct an $N$-step constant trajectory also the complementary slackness and the dual feasibility conditions (\ref{eq:KKT_GNEPslack1}-\ref{eq:KKT_GNEPslack2}) are satisfied as they correspond point-wise in time to their counterparts in~\eqref{eq:steadystateGNEKKT}.

It remains to analyze~\eqref{eq:KKT_GNEPbnd}. 
For case (i), notice that the Lagrangian $\mc L^v(x,u, \lambda^v, \mu^v, \nu^v)$ in~\eqref{eq:Lagrangian} has an additional term
$
 V_\mathrm{f}^v(x_N)$
which using standard KKT analysis leads to the modified  boundary condition 
\begin{equation}\label{eq:KKT_GNEPbndi}
 x_0 =\x \quad\text{and}\quad    \lambda_N = \nabla V_\mathrm{f}^v(x_N) + g_x^\top\mu_N^v. \tag{\ref{eq:KKT_GNEP}d-i}
\end{equation} 
Recall that the assertion sets $\x = x_s$. 
From the condition that $\nabla V_\mathrm{f}^v(x_s) + g_x^\top\mu_N^v = \lambda_s^v$ it then immediately follows that the $N$-step constant trajectory staying at $(x_s, u_s^v, \lambda_s^v, \mu_s^v, \nu_s^v)$ satisfies $x_0=x_s$ and~\eqref{eq:KKT_GNEPbndi}.  

For case (ii), observe that the Lagrangian $\mc L^v$
now includes the additional term
$
 \sigma^\top(x_N-x_s)$,
where $\sigma \in \mathbb R^n$ is the additional multiplier for the point-wise terminal constraint. 
Once more, standard KKT analysis leads to the modified boundary condition
\begin{equation}\label{eq:KKT_GNEPbndii}
 x_0 =\x \quad\text{and}\quad   x_N = x_s. \tag{\ref{eq:KKT_GNEP}d-ii}
\end{equation} 
The $N$-step constant trajectory staying at $(x_s, u_s^v, \lambda_s^v,$  $\mu_s^v, \nu_s^v)$ satisfies these boundary conditions as $\x = x_s$.  

The preceding analysis has shown that the constant trajectory staying at $(x_s, u_s^v, \lambda_s^v, \mu_s^v, \nu_s^v)$  satisfies the KKT conditions~\eqref{eq:KKT_GNEP}. Theorem~\ref{thm:KKTGNEP} then directly gives the assertion. 
This finishes the proof. 
\end{proof}

A first connection between the KKT system and the role of the terminal constraint on the dual variable was drawn in~\cite[Prop. 3]{hall2025stability} and the proof therein. Specifically, it was demonstrated that with a terminal constraint $x_N= x_s$ and additional assumptions on the cost, the steady-state GNEP $(x_s, u_s)$ is the unique fixed point of the finite-horizon GNEP-KKT system and the unique equilibrium point of a closed-loop system resulting from a receding-horizon implementation of the GNE.

Next, we design a linear end penalty which suppresses the leaving arc.
\begin{cor}[Linear end penalty]\label{cor:linEndPen}
Let the conditions of Proposition~\ref{prop:NoLeavArc} (i) hold and suppose that $\mu_s^v = 0$ for all $v\in \mc V$. Then
\[
V_\mathrm{f}^v(x) = x^\top \lambda_s^v, \quad \forall v\in \mc V
\]
gives that, for any horizon $N\in \mathbb N_{>0}$, the constant trajectory $(x, u) \equiv (x_s, u_s)$ satisfies
\[(x_s, u_s)\equiv: (x, u) = (x^*, u^*) \in \mc S_N^{GNE}(x_s). \]
\end{cor}
Indeed, quite similarly to the optimal control setting~\cite{zanon2018economic}, also for GNEPs~\eqref{eq:MPCPerAgent}, the linear end penalty $V_\mathrm{f}^v(x) =x^\top \lambda_s^v$ admits a primal and a dual interpretation:
\begin{itemize}
    \item It directly specifies the boundary or terminal constraint on the dual variable $\lambda^v_N$. Interestingly, the point-wise boundary condition on $\lambda^v_N$ is an optimality condition but it does not imply any reachability requirement of the actual primal state dynamics $x^+ = f(x,u)$.
    \item Via a converse telescopic sum argument $V_\mathrm{f}^v(x) = \lambda_s^\top x$ can be  pulled into the stage cost
    \[
    \tilde \ell^v(x, u^v, u^{-v}) = \ell^v(x,u^v, u^{-v}) + \lambda_s^\top\left(f(x, u^v, u^{-v})-x\right).
    \]
    This is called \textit{rotating the stage cost} in economic MPC~\cite{diehl2011lyapunov,angeli2012average}. The linear rotation with $\lambda_s$ implies that the steady-state constraint in~\eqref{eq:SteadyStateGNEP} is only weakly active. It is also called gradient correction, cf.~\cite{zanon2018economic}. 
\end{itemize}

\subsection{Learning the linear end penalty}
\label{subsec:LearningPenalty}

Both the point-wise terminal constraint $x_N=x_s$ and the linear end penalty approach $V_\mathrm{f}^v(x) = x^\top \lambda_s^v$ require solving the steady-state GNEP~\eqref{eq:SteadyStateGNEP} in advance. Yet, in settings where the cost function changes occasionally (e.g. due to parametric updates such as prices) it would be desirable to avoid the extra computational effort. In the context of suppressing turnpike leaving arcs in OCPs,~\cite[Sec. 4]{faulwasser2018asymptotic} proposed an adaptive strategy for the terminal penalty. 

Algorithm~\ref{alg:learnVf} summarizes how this strategy can be transferred to GNEPs. In Step 2 the GNEP~\eqref{eq:MPCPerAgent} is solved considering the linear end penalty $V_\mathrm{f}^v(x) = x^\top p^v_i$, where the initialization $p_0=0$ is used. 
Then, based on our turnpike insights, for all $v\in \mc V$, an approximation of the turnpike dual variable $\lambda^v_s$ is obtained by $p_i^v = \lambda^v(\frac N 2)$ in Step 5. 
The GNEP is recomputed using the updated end penalty $V_\mathrm{f}^v(x) = x^\top p^v_i$.  The algorithm ends if either the change in $\|p_i^v - p_{i-1}^v\|$ becomes small or an iteration limit is reached. 
Put differently, the algorithm builds upon the observation that whenever the states and inputs are close to their turnpike values also the dual variables stay close to their turnpike values.
It is worth noting that each agent can learn its specific linear end penalty. One may even learn this end penalty for some agent while others still exhibit leaving arcs. 

While the formal convergence analysis of Algorithm~\ref{alg:learnVf} is beyond the scope of this work , the next section will present numerical results demonstrating its efficacy.

\begin{algorithm}[t!]
\caption{Learning the linear end penalty $\lambda_s^v$}
\label{alg:learnVf}
\begin{algorithmic}[1]
    \Require GNEP~\eqref{eq:MPCPerAgent} with horizon $N = 2\hat N$, initial condition $\x$, and $V_\mathrm{f}^v(x) = x^\top p^v_i$, $i=0, p_0 = 0$, $i_{max}\in \mathbb N$, $\Delta_0 = \infty$, $\varepsilon>0$
    \Ensure $\hat\lambda_s^v, v\in \mc V$
    \While {$i\leq i_{max}$ \textbf{AND} $\Delta_i \geq \varepsilon$}
    \State Solve GNEP~\eqref{eq:MPCPerAgent} with $V_\mathrm{f}^v(x) = x^\top p^v_i$
    \State $i \leftarrow i+1$
     \ForAll{$v \in \mc V$}
    \State $p_i^v \leftarrow \lambda^v(\frac N 2)$
    \EndFor
    \State $\Delta_i \leftarrow \sum_{v\in \mc V}\| p^v_i - p^v_{i-1}\|$
    \EndWhile
    \State \Return $\hat\lambda_s^v = p_i^v,v\in \mc V$
\end{algorithmic}
\end{algorithm}



\section{Simulation Study}\label{sec:Sim}

We consider a simple example of the GNEP in~\eqref{eq:MPCPerAgent} in which we have coupled LTI dynamics, coupled costs, and constraints $\forall v \in \mathcal{V}=\{1,2\}:$  
\begin{equation}\label{eq:GNEP_example}
\left\{
\begin{array}{r l}
\displaystyle \min_{u^v, x}  &\displaystyle \sum_{k= 0}^{N-1}  \displaystyle u_k^v(\sum_{j\in \mc{V}} R^{v,j} u_k^{j}) + \|x_k- x^{\text{ref}}\|_{Q^v}^2\\
\textrm{s.t.} &  x_{k+1} = A x_k +  \displaystyle \sum_{j\in \mc{V}} B^j u^j ,  \hspace{0.1em}  k \in \bb{Z}_{N}\\
& -2\leq u_k^v \leq 2, \hspace{4.6em}
k \in \bb{Z}_{N} \\
& -2\leq \displaystyle \sum_{j\in\mc{V}} u_k^j\leq 2, \hspace{2.9em}
k \in \bb{Z}_{N}
\\ 
&  - 1\leq x_k\leq 1,   x_0 =1. \hspace{1.3em} k \in \bb{Z}_{N+1}.
\end{array}
\right.
\end{equation}
The parameter values are $A = 1.5$, $B^1 = 1$, $B^2 = 2$, $R^{1,1} = R^{1,2} = 4$, $R^{2,2} = R^{2,1} = 5$ and state weights $Q^1 = 1, Q^2 = 2$. The reference state is chosen as $x^{\text{ref}}=0.3$. Assumptions~\ref{ass:ContDiff} and~\ref{ass:Convex} are fulfilled by construction.

\begin{figure}[t]
\centering 
\begin{subfigure}{\columnwidth}
\includegraphics[width=\columnwidth]{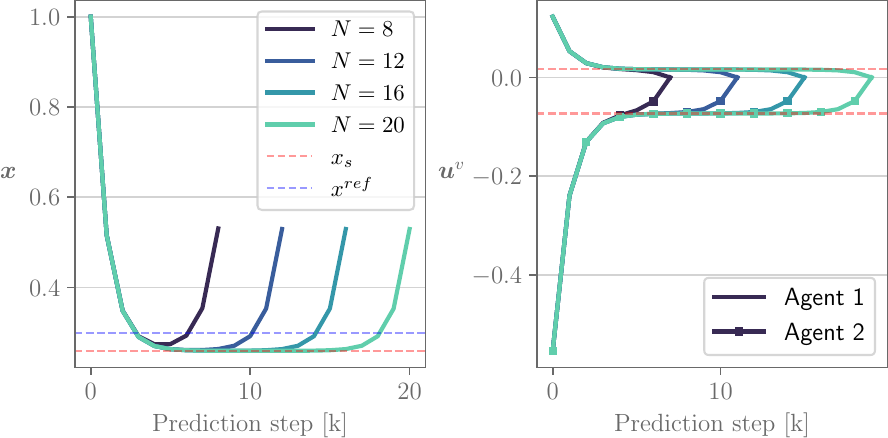}
\caption{}
\label{fig:StateWithoutPenalty}
\end{subfigure}
\begin{subfigure}{0.5\columnwidth}
\includegraphics[width = \columnwidth]{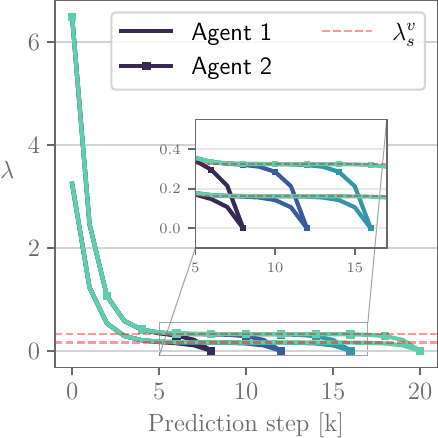}
\caption{}
\label{fig:DualTrajectory}
\end{subfigure}
\caption{Open-loop GNE trajectories of~\eqref{eq:GNEP_example} without terminal penalty.}
\end{figure}

The game pairs $(x^*,u^*)\in  \mc{S}^{\text{\tiny GNE}}_N(\x)$ for different horizon lengths are shown in Figure~\ref{fig:StateWithoutPenalty} and the corresponding trajectory of dual variables $\lambda^{v}$ in Figure~\ref{fig:DualTrajectory}. We clearly see the characteristic turnpike phenomenon of both input and state trajectories converging to the steady-state GNE $(x_s,u_s)$ and diverging in the final time steps.

\begin{figure}[t]
\centering 
\includegraphics[width=\columnwidth]{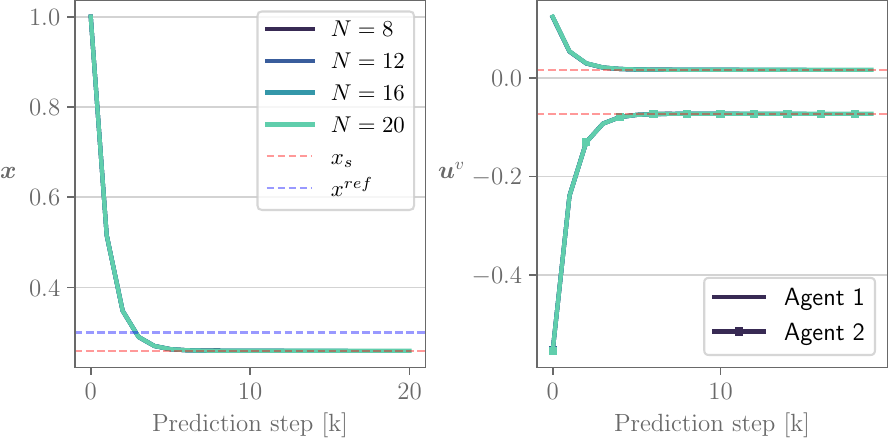}
\caption{Open-loop GNE trajectories of~\eqref{eq:GNEP_example} with a  $\lambda_s\, x_N$ penalty.} \label{fig:StateWithSteadyStatePenalty}
\end{figure}

\begin{figure}[t]
\centering 
\includegraphics[width=\columnwidth]{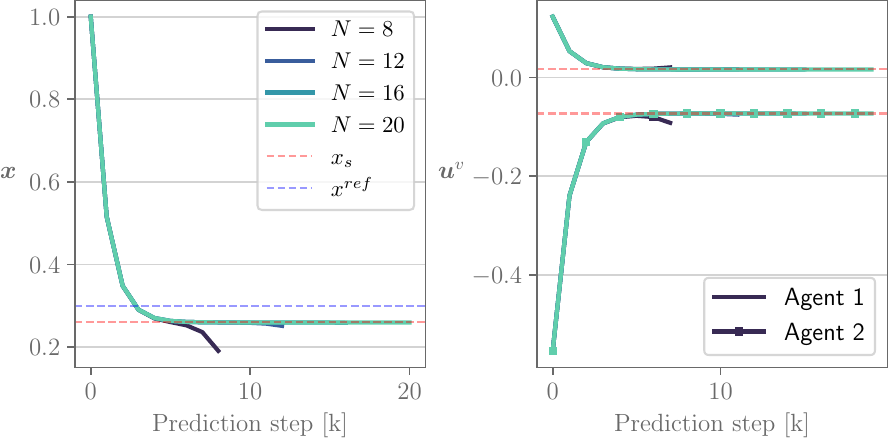}
\caption{Open-loop GNE trajectories of~\eqref{eq:GNEP_example} with a learned terminal penalty following Algorithm~\ref{alg:learnVf} and stopping after one iteration.} \label{fig:StateWithPenaltyLearning}
\end{figure}

As outlined in Section~\ref{subsec:LinearEndPenalty} we apply a terminal penalty to the cost function of each agent which becomes $J^v(x,u^{v},u^{-v}) + (\lambda_s^v)^\top  x_N$ the resulting trajectories are presented in Figure~\ref{fig:StateWithSteadyStatePenalty}. Notice that with the terminal penalty, the state and input trajectories converge to the turnpike and remain there until the end of the horizon. In Figure~\ref{fig:StateWithPenaltyLearning}, we demonstrate that in the very simple GNE problem~\eqref{eq:GNEP_example} the penalty learning Algorithm~\ref{alg:learnVf} suppresses the leaving arc substantially even after just one iteration, i.e., $\lambda^{v}_k$ is close to the steady-state multipliers $\lambda^{v}_s$ at the midpoint of the trajectory.

\begin{rmk} Since $\mc{S}^{\text{\tiny GNE}}_N(\x)$ and the steady-state GNE in~\eqref{eq:SteadyStateGNEP} are not necessarily unique, the terminal penalty method requires verifying that $\lambda_s$ corresponds to the global turnpike. Local turnpike phenomenon in OCPs have been studied in~\cite{kruegel2023local}, but this is left for future work in the GNE setting.
\end{rmk}

\section{Conclusion}

In this manuscript we study GNE problems and derive system-theoretic insights into their open-loop trajectories based on dissipativity analysis. We extend the close relation between dissipativity and turnpike behaviors known in optimal control to dynamic generalized Nash games. The two main ingredients of our analysis are a novel definition of a strict dissipativity property tailored to GNEPs as well as an assumption upper bounding the price of anarchy. Moreover, we establish a local variational characterization of the game value function and relate them to the geometry of storage functions. 

Our simulation results show that the linear end penalties designed based on the optimality conditions successfully suppress the leaving arc of the turnpike even in the coupled, non-cooperative, game-theoretic setting. 

In the future, we will investigate adaptive end penalties which would allow to learn $\lambda_s^v$ by extracting it from open-loop predictions while the game is being played rather than having to solve the steady-state GNE problem in advance. Further, we will study closed-loop stability based on turnpike results when GNEs are applied in a receding-horizon fashion as in MPC.

\bibliographystyle{apalike}        
\bibliography{Dissipativity_Turnpikes_GNEPs}

@Article{diehl2011lyapunov,
  author     = {Moritz Diehl and Rishi Amrit and James B. Rawlings},
  journal    = {{IEEE} Transactions on Automatic Control},
  title      = {{A Lyapunov Function for Economic Optimizing Model Predictive Control}},
  year       = {2011},
  month      = mar,
  number     = {3},
  pages      = {703--707},
  volume     = {56},
  doi        = {10.1109/tac.2010.2101291},
  publisher  = {Institute of Electrical and Electronics Engineers ({IEEE})},
  readstatus = {read},
  relevance  = {relevant},
}

@Article{slade1994what,
  author    = {Slade, Margaret E},
  journal   = {The Journal of Industrial Economics},
  title     = {What does an oligopoly maximize?},
  year      = {1994},
  pages     = {45--61},
  publisher = {JSTOR},
}

@Article{monderer1996potential,
  author    = {Monderer, Dov and Shapley, Lloyd S.},
  journal   = {Games and Economic Behavior},
  title     = {Potential Games},
  year      = {1996},
  issn      = {0899-8256},
  month     = may,
  number    = {1},
  pages     = {124--143},
  volume    = {14},
  doi       = {10.1006/game.1996.0044},
  publisher = {Elsevier BV},
}

@InProceedings{hall2022receding,
  author    = {Hall, Sophie and Belgioioso, Giuseppe and Liao-McPherson, Dominic and Dorfler, Florian},
  booktitle = {2022 IEEE 61st Conference on Decision and Control (CDC)},
  title     = {Receding Horizon Games with Coupling Constraints for Demand-Side Management},
  year      = {2022},
  pages     = {3795-3800},
  doi       = {10.1109/CDC51059.2022.9992497},
}

@Article{vincent2020influence,
  author    = {Vincent, Rémy and Houari, Azeddine and Ait-Ahmed, Mourad and Benkhoris, Mohamed Fouad},
  journal   = {Journal of Energy Storage},
  title     = {Influence of different time horizon-based battery energy management strategies on residential microgrid profitability},
  year      = {2020},
  issn      = {2352-152X},
  month     = jun,
  pages     = {101340},
  volume    = {29},
  doi       = {10.1016/j.est.2020.101340},
  publisher = {Elsevier BV},
}

@InBook{noor2025recent,
  author    = {Noor, Muhammad Aslam and Noor, Khalida Inayat and Rassias, Michael Th.},
  pages     = {315--359},
  publisher = {Springer Nature Switzerland},
  title     = {Recent Developments in General Quasi Variational Inequalities},
  year      = {2025},
  isbn      = {9783031870576},
  booktitle = {Geometry and Non-Convex Optimization},
  doi       = {10.1007/978-3-031-87057-6_13},
  issn      = {1931-6836},
}

@Article{dutang2013existence,
  author      = {Dutang, Christophe},
  journal     = {{Journal of Nonlinear Analysis and Optimization}},
  title       = {{Existence theorems for generalized Nash equilibrium problems: an analysis of assumptions}},
  year        = {2013},
  number      = {2},
  pages       = {115-126},
  volume      = {4},
  hal_id      = {hal-00828948},
  hal_version = {v2},
  publisher   = {{Sompong Dhompongsa and Somyot Plubtieng}},
  url         = {https://hal.science/hal-00828948},
}

@InBook{alphonse2021stability,
  author    = {Alphonse, Amal and Hintermüller, Michael and Rautenberg, Carlos N.},
  pages     = {183--210},
  publisher = {Springer International Publishing},
  title     = {Stability and Sensitivity Analysis for Quasi-Variational Inequalities},
  year      = {2021},
  isbn      = {9783030793937},
  month     = jun,
  booktitle = {Non-Smooth and Complementarity-Based Distributed Parameter Systems},
  doi       = {10.1007/978-3-030-79393-7_8},
  issn      = {2296-6072},
}

@Article{byrnes1994losslessness,
  author   = {Byrnes, C.I. and Wei Lin},
  journal  = {IEEE Transactions on Automatic Control},
  title    = {Losslessness, feedback equivalence, and the global stabilization of discrete-time nonlinear systems},
  year     = {1994},
  number   = {1},
  pages    = {83-98},
  volume   = {39},
  doi      = {10.1109/9.273341},
}

@Article{gruene2016relation,
  author    = {Grüne, Lars and Müller, Matthias A.},
  journal   = {Systems \& Control Letters},
  title     = {On the relation between strict dissipativity and turnpike properties},
  year      = {2016},
  issn      = {0167-6911},
  month     = apr,
  pages     = {45--53},
  volume    = {90},
  doi       = {10.1016/j.sysconle.2016.01.003},
  publisher = {Elsevier BV},
}

@Article{hall2025stability,
  author   = {Hall, Sophie and Belgioioso, Giuseppe and Dörfler, Florian and Liao-McPherson, Dominic},
  journal  = {IEEE Transactions on Automatic Control},
  title    = {Stability Certificates for Receding Horizon Games},
  year     = {2025},
  pages    = {1-8},
  doi      = {10.1109/TAC.2025.3647314},
}

@Article{faulwasser2022turnpike,
  author    = {Faulwasser, Timm and Gr{\"u}ne, Lars},
  journal   = {Handbook of numerical analysis},
  title     = {Turnpike properties in optimal control: An overview of discrete-time and continuous-time results},
  year      = {2022},
  pages     = {367--400},
  volume    = {23},
  publisher = {Elsevier},
}

@Article{gruene2013economic,
  author    = {Lars Grüne},
  journal   = {Automatica},
  title     = {Economic receding horizon control without terminal constraints},
  year      = {2013},
  month     = {mar},
  number    = {3},
  pages     = {725--734},
  volume    = {49},
  doi       = {10.1016/j.automatica.2012.12.003},
  publisher = {Elsevier {BV}},
}

@InCollection{facchinei2009nash,
  author    = {Francisco Facchinei and Jong Pang},
  booktitle = {Convex Optimization in Signal Processing and Communications},
  publisher = {Cambridge University Press},
  title     = {{Nash equilibria: the variational approach}},
  year      = {2009},
  chapter   = {12},
  editor    = {Daniel P. Palomar and Yonina C. Eldar},
  month     = {dec},
  pages     = {443--493},
  doi       = {10.1017/cbo9780511804458.013},
  place     = {Cambridge},
}

@Article{chen2021distributed,
  author    = {Chen, Guanpu and Ming, Yang and Hong, Yiguang and Yi, Peng},
  journal   = {Automatica},
  title     = {{Distributed algorithm for $\varepsilon$-generalized Nash equilibria with uncertain coupled constraints}},
  year      = {2021},
  issn      = {0005-1098},
  month     = jan,
  pages     = {109313},
  volume    = {123},
  doi       = {10.1016/j.automatica.2020.109313},
  publisher = {Elsevier BV},
}

@Article{bueno2019optimality,
  author    = {Bueno, Lu\'{\i}s Felipe and Haeser, Gabriel and Rojas, Frank Navarro},
  journal   = {SIAM Journal on Optimization},
  title     = {Optimality Conditions and Constraint Qualifications for Generalized Nash Equilibrium Problems and Their Practical Implications},
  year      = {2019},
  number    = {1},
  pages     = {31-54},
  volume    = {29},
  doi       = {10.1137/17M1162524},
  eprint    = {https://doi.org/10.1137/17M1162524},
  relevance = {relevant},
  url       = {https://doi.org/10.1137/17M1162524},
}

@Article{facchinei2009generalized,
  author    = {Facchinei, Francisco and Kanzow, Christian},
  journal   = {Annals of Operations Research},
  title     = {Generalized Nash Equilibrium Problems},
  year      = {2009},
  issn      = {1572-9338},
  month     = nov,
  number    = {1},
  pages     = {177--211},
  volume    = {175},
  doi       = {10.1007/s10479-009-0653-x},
  publisher = {Springer Science and Business Media LLC},
}

@Article{angeli2012average,
  author     = {Angeli, David and Amrit, Rishi and Rawlings, James B.},
  journal    = {IEEE Transactions on Automatic Control},
  title      = {On Average Performance and Stability of Economic Model Predictive Control},
  year       = {2012},
  number     = {7},
  pages      = {1615-1626},
  volume     = {57},
  doi        = {10.1109/TAC.2011.2179349},
  readstatus = {read},
}

@Article{mueller2015role,
  author       = {Matthias A. Müller and Lars Grüne and Frank Allgöwer},
  journal      = {{IFAC}-{PapersOnLine}},
  title        = {On the role of dissipativity in economic model predictive control},
  year         = {2015},
  number       = {23},
  pages        = {110--116},
  volume       = {48},
  date         = {2015},
  doi          = {10.1016/j.ifacol.2015.11.269},
  journaltitle = {{IFAC}-{PapersOnLine}},
  publisher    = {Elsevier {BV}},
}

@Article{lecleach2022algames,
  author    = {Le Cleac’h, Simon and Schwager, Mac and Manchester, Zachary},
  journal   = {Autonomous Robots},
  title     = {ALGAMES: a fast augmented Lagrangian solver for constrained dynamic games},
  year      = {2022},
  issn      = {1573-7527},
  number    = {1},
  pages     = {201--215},
  volume    = {46},
  doi       = {10.1007/s10514-021-10024-7},
  publisher = {Springer},
  refid     = {Le Cleac’h2022},
  url       = {https://doi.org/10.1007/s10514-021-10024-7},
}

@InBook{carlson1995turnpike,
  author    = {Carlson, D. and Haurie, A.},
  pages     = {353--376},
  publisher = {Birkhäuser Boston},
  title     = {A Turnpike Theory for Infinite Horizon Open-Loop Differential Games with Decoupled Controls},
  year      = {1995},
  isbn      = {9781461242741},
  booktitle = {New Trends in Dynamic Games and Applications},
  doi       = {10.1007/978-1-4612-4274-1_18},
}

@Article{carlson1996turnpike,
  author    = {Carlson, D. and Haurie, A.},
  journal   = {SIAM Journal on Control and Optimization},
  title     = {A Turnpike Theory for Infinite-Horizon Open-Loop Competitive Processes},
  year      = {1996},
  issn      = {1095-7138},
  month     = jul,
  number    = {4},
  pages     = {1405--1419},
  volume    = {34},
  doi       = {10.1137/s0363012994265158},
  publisher = {Society for Industrial & Applied Mathematics (SIAM)},
}

@InBook{carlson2000infinite,
  author    = {Carlson, Dean A. and Haurie, Alain B.},
  pages     = {195--212},
  publisher = {Birkhäuser Boston},
  title     = {Infinite Horizon Dynamic Games with Coupled State Constraints},
  year      = {2000},
  isbn      = {9781461213369},
  booktitle = {Advances in Dynamic Games and Applications},
  doi       = {10.1007/978-1-4612-1336-9_10},
}

@InProceedings{kruegel2023local,
  author    = {Krügel, Lisa and Faulwasser, Timm and Grüne, Lars},
  booktitle = {2023 62nd IEEE Conference on Decision and Control (CDC)},
  title     = {Local Turnpike Properties in Finite Horizon Optimal Control},
  year      = {2023},
  pages     = {5273-5278},
  doi       = {10.1109/CDC49753.2023.10383850},
}

@Article{faulwasser2018asymptotic,
  author    = {Faulwasser, Timm and Zanon, Mario},
  journal   = {IFAC-PapersOnLine},
  title     = {Asymptotic Stability of Economic NMPC: The Importance of Adjoints},
  year      = {2018},
  issn      = {2405-8963},
  number    = {20},
  pages     = {157--168},
  volume    = {51},
  doi       = {10.1016/j.ifacol.2018.11.009},
  publisher = {Elsevier BV},
}

@Article{zanon2018economic,
  author    = {Zanon, Mario and Faulwasser, Timm},
  journal   = {Journal of Process Control},
  title     = {Economic MPC without terminal constraints: Gradient-correcting end penalties enforce asymptotic stability},
  year      = {2018},
  issn      = {0959-1524},
  month     = mar,
  pages     = {1--14},
  volume    = {63},
  doi       = {10.1016/j.jprocont.2017.12.005},
  publisher = {Elsevier BV},
}

@Article{hall2024receding,
  author    = {Hall, Sophie and Guerrini, Laura and Dörfler, Florian and Liao-McPherson, Dominic},
  journal   = {IFAC-PapersOnLine},
  title     = {Receding Horizon Games for Modeling Competitive Supply Chains},
  year      = {2024},
  issn      = {2405-8963},
  number    = {18},
  pages     = {8--14},
  volume    = {58},
  doi       = {10.1016/j.ifacol.2024.09.002},
  publisher = {Elsevier BV},
}

@Misc{kulkarni2019efficiency,
  author    = {Kulkarni, Ankur A.},
  title     = {The Efficiency of Generalized Nash and Variational Equilibria},
  year      = {2019},
  copyright = {arXiv.org perpetual, non-exclusive license},
  doi       = {10.48550/ARXIV.1908.00702},
  publisher = {arXiv},
  relevance = {relevant},
}

@Article{kulkarni2012variational,
  author     = {Ankur A. Kulkarni and Uday V. Shanbhag},
  journal    = {Automatica},
  title      = {On the variational equilibrium as a refinement of the generalized {N}ash equilibrium},
  year       = {2012},
  issn       = {0005-1098},
  number     = {1},
  pages      = {45-55},
  volume     = {48},
  doi        = {https://doi.org/10.1016/j.automatica.2011.09.042},
  readstatus = {skimmed},
  relevance  = {relevant},
  url        = {https://www.sciencedirect.com/science/article/pii/S0005109811004821},
}

@InProceedings{hall2025limits,
  author    = {Hall, Sophie and Dörfler, Florian and Nax, Heinrich H. and Bolognani, Saverio},
  booktitle = {2025 IEEE 64th Conference on Decision and Control (CDC)},
  title     = {The Limits of ``Fairness" of the Variational Generalized Nash Equilibrium},
  year      = {2025},
  pages     = {5354-5360},
  doi       = {10.1109/CDC57313.2025.11312061},
}

@Article{benenati2025linear,
  author   = {Benenati, Emilio and Grammatico, Sergio},
  journal  = {IEEE Transactions on Automatic Control},
  title    = {Linear-Quadratic Dynamic Games as Receding-Horizon Variational Inequalities},
  year     = {2025},
  pages    = {1-16},
  doi      = {10.1109/TAC.2025.3632150},
}

@Article{atzeni2013demand,
  author     = {Atzeni, Italo and Ord{\'o}{\~n}ez, Luis G and Scutari, Gesualdo and Palomar, Daniel P and Fonollosa, Javier Rodr{\'\i}guez},
  journal    = {IEEE Transactions on Smart Grid},
  title      = {Demand-side management via distributed energy generation and storage optimization},
  year       = {2013},
  month      = jun,
  number     = {2},
  pages      = {866-876},
  volume     = {4},
  doi        = {10.1109/TSG.2012.2206060},
  publisher  = {Institute of Electrical and Electronics Engineers ({IEEE})},
  readstatus = {read},
}

@Book{pavel2012game,
  author    = {Pavel, Lacra},
  publisher = {Birkhäuser Boston},
  title     = {Game Theory for Control of Optical Networks},
  year      = {2012},
  isbn      = {9780817683221},
  doi       = {10.1007/978-0-8176-8322-1},
  issn      = {2363-8524},
  journal   = {Static &amp; Dynamic Game Theory: Foundations &amp; Applications},
}

@InBook{bassanini2002allocation,
  author    = {Bassanini, A. and La Bella, A. and Nastasi, A.},
  pages     = {1--17},
  publisher = {Springer US},
  title     = {Allocation of Railroad Capacity Under Competition: A Game Theoretic Approach to Track time Pricing},
  year      = {2002},
  isbn      = {9781475768718},
  booktitle = {Transportation and Network Analysis: Current Trends},
  doi       = {10.1007/978-1-4757-6871-8_1},
  issn      = {1384-6485},
}

@Article{dreves2018generalized,
  author    = {Axel Dreves and Matthias Gerdts},
  journal   = {Optimal Control Applications and Methods},
  title     = {A generalized {N}ash equilibrium approach for optimal control problems of autonomous cars},
  year      = {2018},
  month     = jul,
  number    = {1},
  pages     = {326-342},
  volume    = {39},
  doi       = {https://doi.org/10.1002/oca.2348},
  eprint    = {https://onlinelibrary.wiley.com/doi/pdf/10.1002/oca.2348},
  publisher = {Wiley},
  url       = {https://onlinelibrary.wiley.com/doi/abs/10.1002/oca.2348},
}

@Article{nabetani2009parametrized,
  author     = {Koichi Nabetani and Paul Tseng and Masao Fukushima},
  journal    = {Computational Optimization and Applications},
  title      = {Parametrized variational inequality approaches to~generalized Nash equilibrium problems with~shared constraints},
  year       = {2009},
  month      = may,
  number     = {3},
  pages      = {423--452},
  volume     = {48},
  doi        = {10.1007/s10589-009-9256-3},
  publisher  = {Springer Science and Business Media {LLC}},
  readstatus = {skimmed},
}

@Article{belgioioso2022distributed,
  author       = {Belgioioso, Giuseppe and Yi, Peng and Grammatico, Sergio and Pavel, Lacra},
  journal      = {{IEEE} Control Systems Magazine},
  title        = {{Distributed generalized Nash equilibrium seeking: An operator-theoretic perspective}},
  year         = {2022},
  month        = {aug},
  number       = {4},
  pages        = {87--102},
  volume       = {42},
  date         = {2022},
  doi          = {10.1109/MCS.2022.3171480},
  journaltitle = {IEEE Control Systems Magazine},
  publisher    = {IEEE},
}

@Article{li2025turnpike,
  author        = {Li, Xun and Wu, Fan and Zhang, Xin},
  title         = {Turnpike properties for zero-sum stochastic linear quadratic differential games of Markovian regime switching system},
  year          = {2025},
  month         = sep,
  archiveprefix = {arXiv},
  copyright     = {arXiv.org perpetual, non-exclusive license},
  doi           = {10.48550/ARXIV.2509.09358},
  eprint        = {2509.09358},
  primaryclass  = {math.OC},
  publisher     = {arXiv},
}

@Article{cohen2025turnpike,
  author        = {Cohen, Asaf and Jian, Jiamin},
  title         = {Turnpike properties in linear quadratic Gaussian N-player differential games},
  year          = {2025},
  month         = jul,
  archiveprefix = {arXiv},
  copyright     = {Creative Commons Attribution 4.0 International},
  doi           = {10.48550/ARXIV.2507.11632},
  eprint        = {2507.11632},
  primaryclass  = {math.OC},
  publisher     = {arXiv},
}

@Article{ersland2025long,
  author        = {Ersland, Olav and Jakobsen, Espen Robstad and Porretta, Alessio},
  title         = {Long time behaviour of Mean Field Games with fractional diffusion},
  year          = {2025},
  month         = may,
  archiveprefix = {arXiv},
  copyright     = {Creative Commons Attribution 4.0 International},
  doi           = {10.48550/ARXIV.2505.06183},
  eprint        = {2505.06183},
  primaryclass  = {math.AP},
  publisher     = {arXiv},
}

@Article{fedorov2025studying,
  author    = {Fedorov, F. A.},
  journal   = {Moscow University Computational Mathematics and Cybernetics},
  title     = {Studying the Well-Posedness of the Boundary Value Problem for a System of Riccati Type Equations Based on the Concept of Mean Field Games},
  year      = {2025},
  issn      = {1934-8428},
  month     = jun,
  number    = {2},
  pages     = {150--164},
  volume    = {49},
  doi       = {10.3103/s0278641925700086},
  publisher = {Allerton Press},
}

@Article{cirant2021long,
  author    = {Cirant, Marco and Porretta, Alessio},
  journal   = {ESAIM: Control, Optimisation and Calculus of Variations},
  title     = {Long time behavior and turnpike solutions in mildly non-monotone mean field games},
  year      = {2021},
  issn      = {1262-3377},
  pages     = {86},
  volume    = {27},
  doi       = {10.1051/cocv/2021077},
  editor    = {Buttazzo, G. and Casas, E. and de Teresa, L. and Glowinski, R. and Leugering, G. and Trélat, E. and Zhang, X.},
  publisher = {EDP Sciences},
}

@Article{carmona2024leveraging,
  author    = {Carmona, René A. and Zeng, Claire},
  journal   = {IEEE Open Journal of Control Systems},
  title     = {Leveraging the Turnpike Effect for Mean Field Games Numerics},
  year      = {2024},
  issn      = {2694-085X},
  pages     = {389--404},
  volume    = {3},
  doi       = {10.1109/ojcsys.2024.3419642},
  publisher = {Institute of Electrical and Electronics Engineers (IEEE)},
}

@Article{fershtman1986turnpike,
  author    = {Fershtman, Chaim and Muller, Eitan},
  journal   = {Journal of Economic Theory},
  title     = {Turnpike properties of capital accumulation games},
  year      = {1986},
  issn      = {0022-0531},
  month     = feb,
  number    = {1},
  pages     = {167--177},
  volume    = {38},
  doi       = {10.1016/0022-0531(86)90094-3},
  publisher = {Elsevier BV},
}

@InCollection{vonNeumann38,
  author    = {von Neumann, J.},
  title     = {{\"U}ber ein \"o{}konomisches {G}leichungssystem und eine {V}erallgemeinerung des {B}rouwerschen {F}ixpunktsatzes},
  booktitle = {{E}rgebnisse eines {M}athematischen {S}eminars},
  year      = {1938},
  editor    = {Menger, K.},
  location  = {Vienna, Austria},
}

@Book{Dorfman58,
  Title                    = {Linear Programming and Economic Analysis},
  Author                   = {Dorfman, R. and Samuelson, P.A. and Solow, R.M.},
  Publisher                = {McGraw-Hill, New York},
  Year                     = {1958}
}

@Article{Mckenzie76,
  Title                    = {Turnpike theory},
  Author                   = {McKenzie, L.W.},
  Journal                  = {Econometrica: Journal of the Econometric Society},
  Year                     = {1976},
  Number                   = {5},
  Pages                    = {841--865},
  Volume                   = {44},

  Publisher                = {JSTOR}
}

@Article{Bellman54a,
  author    = {Bellman, R.},
  title     = {The theory of dynamic programming},
  number    = {6},
  pages     = {503--515},
  volume    = {60},
  journal   = {Bulletin of the American Mathematical Society},
  publisher = {American Mathematical Society},
  year      = {1954},
}

@Article{Ramsey28,
  Title                    = {A mathematical theory of saving},
  Author                   = {Ramsey, F. P.},
  Journal                  = {The Economic Journal},
  Year                     = {1928},
  Number                   = {152},
  Pages                    = {543--559},
  Volume                   = {38},

  Publisher                = {JSTOR}
}

@Article{epfl:faulwasser15h,
  author  = {Faulwasser, T. and Korda, M. and Jones, C.N. and Bonvin, D.},
  journal = {Automatica},
  title   = {On Turnpike and Dissipativity Properties of Continuous-Time Optimal Control Problems},
  year    = {2017},
  pages   = {297-304},
  volume  = {81},
  doi     = {10.1016/j.automatica.2017.03.012},
}

@Article{Willems71a,
  author    = {Willems, J.C.},
  title     = {Least squares stationary optimal control and the algebraic Riccati equation},
  number    = {6},
  pages     = {621--634},
  volume    = {16},
  journal   = {IEEE Trans. Autom. Contr.},
  publisher = {IEEE},
  year      = {1971},
}

@Article{Trelat15a,
  Title                    = {The turnpike property in finite-dimensional nonlinear optimal control},
  Author                   = {Tr{\'e}lat, E. and Zuazua, E.},
  Journal                  = {Journal of Differential Equations},
  Year                     = {2015},
  Number                   = {1},
  Pages                    = {81--114},
  Volume                   = {258}
}

@Article{Kellett14,
  Title                    = {A compendium of comparison function results},
  Author                   = {Kellett, C.M.},
  Journal                  = {Mathematics of Control, Signals, and Systems},
  Year                     = {2014},
  Number                   = {3},
  Pages                    = {339--374},
  Volume                   = {26},

  Publisher                = {Springer}
}

@Article{tudo:faulwasser21a,
  author       = {Faulwasser, T. and Kellett, C.M.},
  date         = {2021},
  journaltitle = {Automatica},
  title        = {On Continuous-Time Infinite Horizon Optimal Control -- {D}issipativity, Stability and Transversality},
  doi          = {10.1016/j.automatica.2021.109907},
  pages        = {109907},
  volume       = {134},
  journal      = {Automatica},
  year         = {2021},
}

@Article{Willems72a,
  author    = {Willems, J.C.},
  title     = {Dissipative dynamical systems part I: General theory},
  number    = {5},
  pages     = {321--351},
  volume    = {45},
  journal   = {Archive for Rational Mechanics and Analysis},
  publisher = {Springer},
  year      = {1972},
}

@Article{Stieler14a,
  Title                    = {An exponential turnpike theorem for dissipative optimal control problems},
  Author                   = {Damm, T. and Gr{\"u}ne, L. and Stieler, M. and Worthmann, K.},
  Journal                  = {SIAM Journal on Control and Optimization},
  Year                     = {2014},
  Number                   = {3},
  Pages                    = {1935-1957},
  Volume                   = {52}
}

@Book{palomar2010convex,
  author    = {Daniel P. Palomar and Yonina C. Eldar.},
  publisher = {CAMBRIDGE},
  title     = {Convex Optimization in Signal Processing and Communications},
  year      = {2010},
  isbn      = {0521762227},
  month     = dec,
  volume    = {27},
  doi       = {10.1017/CBO9780511804458},
  ean       = {9780521762229},
  issn      = {1053-5888},
  journal   = {IEEE Signal Processing Magazine},
  pages     = {19-145},
  pagetotal = {498},
  url       = {https://www.ebook.de/de/product/9549359/convex_optimization_in_signal_processing_and_communications.html},
}

@Article{gadjov2019passivity,
  author       = {Gadjov, Dian and Pavel, Lacra},
  journal      = {{IEEE} Transactions on Automatic Control},
  title        = {A Passivity-Based Approach to {N}ash Equilibrium Seeking Over Networks},
  year         = {2019},
  month        = {mar},
  number       = {3},
  pages        = {1077--1092},
  volume       = {64},
  date         = {2019},
  doi          = {10.1109/tac.2018.2833140},
  journaltitle = {IEEE Transactions on Automatic Control},
  publisher    = {Institute of Electrical and Electronics Engineers ({IEEE})},
}

@Article{Willems07a,
  author    = {Willems, J.C.},
  title     = {Dissipative dynamical systems},
  number    = {2-3},
  pages     = {134--151},
  volume    = {13},
  journal   = {European Journal of Control},
  publisher = {Elsevier},
  year      = {2007},
}

@Article{Anderson87a,
  Title                    = {Optimal control problems over large time intervals},
  Author                   = {Anderson, B.D.O. and Kokotovic, P.V.},
  Journal                  = {Automatica},
  Year                     = {1987},
  Number                   = {3},
  Pages                    = {355--363},
  Volume                   = {23},

  Publisher                = {Elsevier}
}

\appendix
\section{Proof of Theorem~\ref{thm:AvailStor}}\label{apendix:ProofThmAvailStor}

We adapt the original proof of \cite{Willems72a} to the setting of dynamic GNEPs and to the strict dissipation inequality \eqref{eq:GsDI}. Henceforth, we use the shorthand $\alpha_\ell(\cdot):= \alpha_{\ell}\left(\left\| \begin{smallmatrix}x_k- x_s\\ u_k - u_s \end{smallmatrix}\right\|\right)$. Where for simplicity we dropped the $.^*$ notation.

$\Lambda_{\alpha_\ell} < \infty \Rightarrow \eqref{eq:GsDI}$: \quad Note that $N=0$ is a feasible but not necessarily optimal solution to \eqref{eq:AvailStorDef} and hence $\Lambda_{\alpha_\ell}(\x) \geq 0$. Consider some trajectory $(x, u) \in \mc{S}^{\text{\tiny GNE}}_N(\x)$ which connects $x_0 = \x$ with $x_m$ in $m \geq 1$ steps. 
One suboptimal possibility to extract the storage at $x_m$ is to first use $(x, u)$ to traverse from $\x$ to $x_m$ and then apply the inputs suggested by \eqref{eq:AvailStorDef}. This would extract the storage equivalent to 
$
\sum_{k=0}^m -  \alpha_{\ell}\left(\cdot\right) +s(x_k,u_k) + \Lambda_{\alpha_\ell}(x_m)$.
The definition of the available storage \eqref{eq:AvailStorDef}, however, requires to find the supremum of
$\sum_{k=0}^N \alpha_\ell(\cdot) - s(x_k, u_k)$  
with free end time $N$. If at $\x = x_0$, the optimal horizon in \eqref{eq:AvailStorDef} $N^*$ is larger than $m$, then $\Lambda_{\alpha_\ell}$ being the optimal value of \eqref{eq:AvailStorDef} gives
\begin{equation}\label{eq:availStorPf1}
    \Lambda_{\alpha_\ell}(\x) \geq \sum_{k=0}^m   \alpha_{\ell}\left(\cdot \right) -s(x_k,u_k) + \Lambda_{\alpha_\ell}(x_m).
\end{equation}
If at $\x = x_0$, we have $N^* < m$, optimality likewise implies the above inequality. Rearranging \eqref{eq:availStorPf1} directly gives \eqref{eq:GsDI}.

$\eqref{eq:GsDI} \Rightarrow \Lambda_a < \infty$:\quad
For the sake of contradiction, suppose that at some $\x$ $\eqref{eq:GsDI}$ holds with storage $\Lambda: \bb X_N(\bb{X}_0)\to \R$ bounded from below while $\Lambda_{\alpha_\ell}(\x) = \infty$.
If \eqref{eq:GsDI} holds for all $(x, u) \in \mc{S}^{\text{\tiny GNE}}_N(\x)$, then at $x_0 = \x$ and for any storage function $\Lambda$ we have
\[
\Lambda(\x)+ \sum_{k=0}^m -  \alpha_{\ell}\left(\cdot\right) +s(x_k,u_k) \geq \Lambda(x_m) \geq -c >-\infty.
\]
The right hand side inequalities follow from the lower boundedness of  storage functions $\Lambda$. Notice that these bounds hold for all $(x, u) \in \mc{S}^{\text{\tiny GNE}}_N(\x)$. 
Hence, we set $m=N$ and rearrange the inequalities  such that
\[\Lambda(\x) \geq -c + \sup_{\substack{N\in \bb{N}\\ (x,u)\\  \in\mc{S}^{\text{\tiny GNE}}_N(\x)}} - \sum_{k=0}^N -  \alpha_{\ell}\left(\cdot\right) +s(x_k,u_k).\]
Here we have used that $-c$ is not affected by the optimization over the free end time $N$.
With $x_0 = \x$, the previous inequality is equivalent to
$
\Lambda(\x) \geq -c + \Lambda_{\alpha_\ell}(\x)$.
Since $c \in \mathbb R$, $\Lambda_{\alpha_\ell}(\x) =\infty$ would contradict $\Lambda(\x)\in \mathbb R$. This finishes the proof.\qed
\section{Performance bounds from bounded PoA}
\label{app:bnd}

\begin{lem}\label{lem:linearBound}
Suppose that $0 < \nu \leq V^\diamond_N(\mathbf{x}) \leq V< \infty$ is satisfied. 
\begin{itemize}
    \item[(i)]If Assumption~\ref{ass:PoA} holds with $P\in \R$, then 
 \begin{equation}\label{eq:linearBound}
     \sup_{(x^*, u^*) \in \mc{S}^{\text{\tiny GNE}}_N(\x)} J_N(x^*, u^*) - V^\diamond_N(\mathbf{x}) \leq VP.
 \end{equation} 
 \item[(ii)] If \eqref{eq:linearBound} holds with bound $VP$, then Assumption~\ref{ass:PoA} holds with $\frac{V}{\nu}P +1$.
\end{itemize}
\end{lem}
\begin{proof}
Part (i): \quad If Assumption~\ref{ass:PoA} holds, we have
\[
\dfrac{1}{ V^\diamond_N(\mathbf{x})}\displaystyle\sup_{(x^*, u^*) \in \mc{S}^{\text{\tiny GNE}}_N(\x)} J_N(x^*, u^*) \leq P \leq P+1.
\]
Multiplying both sides with $V^\diamond_N(\mathbf{x})\geq \nu>0$ gives
\[\sup_{(x^*, u^*) \in \mc{S}^{\text{\tiny GNE}}_N(\x)} J_N(x^*, u^*) \leq V^\diamond_N(\mathbf{x}) (P+1).\]
Substracting $V^\diamond_N(\mathbf{x})$ and using $P\cdot V^\diamond_N(\mathbf{x}) \leq P V$
gives the assertion. 
Part (ii): \quad Consider \eqref{eq:linearBound}, add $V^\diamond_N(\mathbf{x})$ on both sides, and divide by $V^\diamond_N(\mathbf{x})$ to obtain
\[
\dfrac{1}{ V^\diamond_N(\mathbf{x})}\displaystyle\sup_{(x^*, u^*) \in \mc{S}^{\text{\tiny GNE}}_N(\x)} J_N(x^*, u^*) \leq \dfrac{V}{V^\diamond_N(\mathbf{x})}P + 1
\]
The left hand side is bounded from above by $\frac{V}{\nu}P+1$ which finishes the proof. 
\end{proof}
\color{black}
\end{document}